%% file: corona_tracking.tex
\documentclass[journal]{vgtc}
\usepackage{amssymb,amsmath}
\usepackage{xcolor}
\usepackage{hyperref}
\usepackage{graphicx}
\usepackage{subcaption}
\usepackage{booktabs}
\usepackage[square, sort, numbers]{natbib}

\usepackage{enumitem}
\setlist[enumerate]{topsep=1pt}
\providecommand{\tightlist}{%
  \setlength{\itemsep}{0pt}\setlength{\parskip}{0pt}}

\graphicspath{{./plots/}}

\onlineid{0}

\vgtccategory{Research}
\vgtcpapertype{system}

\title{NewsStand CoronaViz: A Map Query Interface for Spatio-Temporal and Spatio-Textual Monitoring of Disease Spread\thanks{This work was sponsored in part by the NSF under Grant iis-18-16889.}}

\author{John Kastner, Hanan Samet, Hong Wei}

\authorfooter{
\item
 John Kastner with University of Maryland. E-mail: kastner@umd.edu.
\item
 Hanan Samet with University of Maryland. E-mail: hjs@cs.umd.edu.
\item
 Hong Wei with University of Maryland. E-mail: hyw@cs.umd.edu.
}

\abstract{
With the rapid continuing spread of COVID-19, it is clearly important to be
able to track the progress of the virus over time in order to be better
prepared to anticipate its emergence and spread in new regions as well as
declines in its presence in regions thereby leading to or justifying
``reopening'' decisions.  There are many applications and web sites that
monitor officially released numbers of cases which are likely to be the most
accurate methods for tracking the progress of the virus; however, they will not
necessarily paint a complete picture.  To begin filling any gaps in official
reports, we have developed the NewsStand CoronaViz web application
(\url{https://coronaviz.umiacs.io}) that can run on desktops and mobile devices that
allows users to explore the geographic spread in discussions about the virus
through analysis of keyword prevalence in geotagged news articles and tweets in
relation to the real spread of the virus as measured by confirmed case numbers
reported by the appropriate authorities.

NewsStand CoronaViz users have access to dynamic variants of the
disease-related variables corresponding to the numbers of confirmed
cases, active cases, deaths, and recoveries (where they are provided)
via a map query interface.  It has the ability to step forward and
backward in time using both a variety of temporal window sizes (day,
week, month, or combinations thereof) in addition to user-defined
varying spatial window sizes specified by direct manipulation actions
(e.g., pan, zoom, and hover) as well as textually (e.g., by the name
of the containing country, state or province, or county as well as
textually-specified spatially-adjacent combinations thereof), and
finally by the amount of spatio-temporally-varying news and tweet
volume involving COVID-19.}

\teaser{
  \centering
  \includegraphics[width=.90\columnwidth]{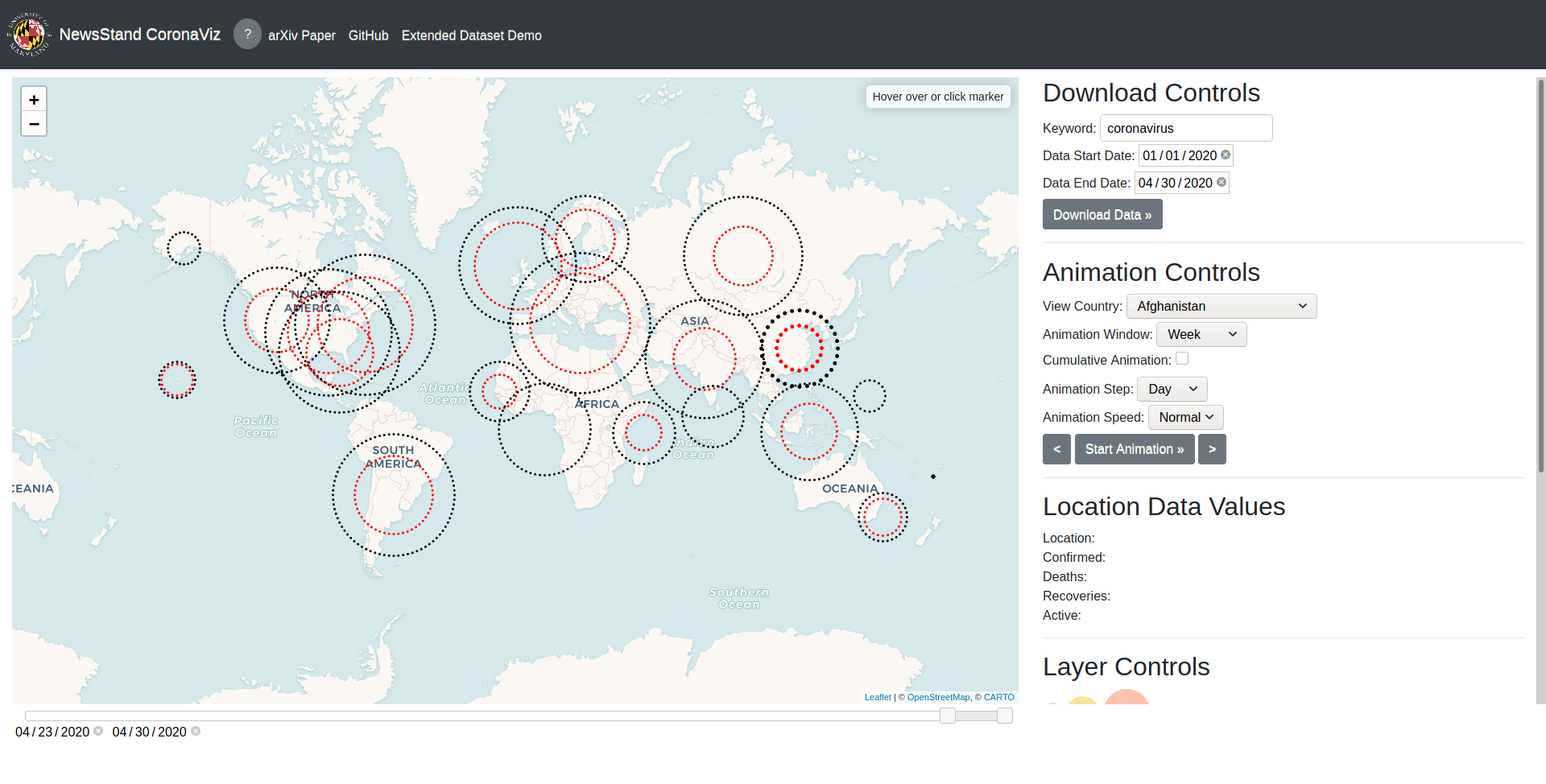}
  \caption{NewsStand CoronaViz application screenshot displaying the confirmed cases and deaths
  layers for a one week period between April 23 and April 30, 2020.
  \label{fig:screenshot}} }

\vgtcinsertpkg

\begin{document}

\firstsection{Introduction}

\maketitle

With the rapid continuing spread of COVID-19, it is clearly important to be
able to track the progress of the virus over time in order to be better
prepared to anticipate its emergence and spread in new regions as well as
declines in its presence in regions thereby leading to or justifying
``reopening'' decisions.  There are many applications and web sites that
monitor officially released numbers of cases which are likely to be the most
accurate methods for tracking the progress of the virus; however, they will not
necessarily paint a complete picture.  To begin filling any gaps in official
reports, we have developed the NewsStand CoronaViz (abreviated as CoronaViz)
web application that can run on desktops and mobile devices that allows users
to explore the geographic spread in discussions about the virus through
analysis of keyword prevalence in geotagged news articles and tweets in
relation to the real spread of the virus as measured by confirmed case numbers
reported by the appropriate authorities.

CoronaViz users have access to dynamic variants of the
disease-related variables corresponding to the numbers of confirmed
cases, active cases, deaths, and recoveries (where they are provided)
via a map query interface.  It has the ability to step forward and
backward in time using both a variety of temporal window sizes (day,
week, month, or combinations thereof) in addition to user-defined
varying spatial window sizes specified by direct manipulation actions
(e.g., pan, zoom, and hover) as well as textually (e.g., by the name
of the containing country, state or province, or county as well as
textually-specified spatially-adjacent combinations thereof), and
finally by the amount of spatio-temporally -varying news and tweet
volume involving COVID-19.

Any combination of the
variables can be viewed subject to a possibility of clutter which is
avoided by the use of concentric circles (termed geo-circles whose
radii correspond to ranges of variable values. The variable values are
provided both on cumulative and day-by-day basis. The visualization
enables spatial, temporal, and keyword variation (i.e., it can be used
for names of other disease names or entirely other concepts such as
names of brands, people, etc. with an appropriate set of variables and
document collections).

It was motivated by the continuing spread of COVID-19 which led to the
desire to track its progress over time to be better prepared to
anticipate its emergence in new regions. There exist numerous systems
to monitor and map officially released numbers of cases~\cite{dong2020interactive} which are
the current established means of keeping track of the progress of the
virus. However, these systems do not necessarily paint a complete
picture. For example, they are primarily mashups in that they do not
support zooming in on the map in the sense that they just increase the
resolution of the map but do not show the data for the additional
units (e.g., states/provinces, counties, etc.) that have become
visible as a result of the zoom.

NewsStand CoronaViz is designed to fill in gaps in the official
reports thereby providing a more complete picture. It incorporates the
NewsStand system~\cite{Teit08,Same14} (see also the
related TwitterStand system~\cite{Sank09c})
which are example applications of a general framework being developed
at the University of Maryland at College Park
to enable searching for information using a map query interface. When
the information domain is news, the underlying search domain results
from monitoring the output of over 10,000 RSS news sources and is
available for retrieval within minutes of publication. The advantage
of doing so is that a map, coupled with the ability to vary the zoom
level at which it is viewed, provides an inherent granularity to the
search process that facilitates an approximate search.

NewsStand CoronaViz makes use of NewsStand to find all news articles
and tweets (identified by containing a pointer to a URL of an RSS
feed) that contain the keyword (COVID-19 or Coronavirus in our
case). It also identifies each toponym (geographic location) that is
mentioned in the article or tweet. Next, it takes the cross product of
these sets as the set of geocoded keywords. In other words, a pair
associating every keyword in the article or tweet with every location
mentioned in the article or tweet. Each of the keyword location pairs
is also associated with the time of publication of the article in
order to enable the temporal component of NewsStand CoronaViz. The
result is the ability to explore the spread of the disease through
analysis of keyword prevalence in geotagged news articles and tweets
over spatial and temporal ranges.

NewsStand CoronaViz is motivated by our previous efforts in devising the
STEWARD~\cite{Lieb07}, NewsStand~\cite{Teit08,Same14}, and
TwitterStand~\cite{Sank09c} systems which made use of a map query interface for
accessing HUD research documents, news, and tweets, respectively, that
exploited the power of spatial synonyms so that users need not know the exact
name of the place for which they are seeking news articles and tweets that
contain (i.e., mention) non-spatial information such as a disease name or any
instance of a concept for which we have an ontology (e.g., names of companies
or brands~\cite{Abde15}, drugs and clinical trials involving
them~\cite{Adel10}, as well as, for example, crimes~\cite{Waji16}).  Instead
they obtain the location via the direct manipulation actions of pan and zoom.
The NewsStand system was adapted for tracking mentions of disease names in
documents~\cite{Lan12} and news articles~\cite{Lan14} but not in conjunction
with other spatio-temporally varying variables such as confirmed cases, deaths,
etc. as we do here.

The rest of this paper is organized as follows. Section \ref{queries}
discusses the queries our system is able to support. Section \ref{relatedwork}
reviews related work in spatiotemporal visualization as well as existing
disease monitoring systems.  Section~\ref{data-collection} describes our data
collections procedures for news articles and tweets. Section~\ref{application-interface}
contains a description of our application interface. Section~\ref{evaluation} is
an evaluation of the utility of news article keyword trends for tracking of diseases spread.
Finally, in Section~\ref{conclusion}, we concluded and remark on possible
directions for future work.

\section{Queries} \label{queries}

In this section we review the queries that NewsStand CoronaViz was
designed to execute.  It is supported by a full-blown database
management system designed to execute queries rapidly often making
use of indexes.  Most of the queries are converted to database queries.
It was originally designed to deal with a spatiotextual database
and has now been augmented by a spatiotemporal component.

The values of all of the variables in NewsStand CoronaViz are
presented in a time-varying manner as time moves on with the aid
of a time slider thereby leading them to be characterized as
{\em dynamic} variables.  This is in contrast to 
visualization tools where such variables are presented in a graph
where time is the horizontal axis and the variable value is the
vertical axis thereby leading them to be characterized as
{\em static}.  Thus we see that the presentation manner is the
key to the characterization.
It is not easy to present several static variables as they tend
to clutter the display regardless of whether they are represented
as one graph for the set of all variables or one graph per variable.
The situation becomes more complex when values of the variables vary in
a spatially-varying manner.  In this case the only way to deal with
the static variables is to repeat the graph at each location.
This is OK when the data is spatially sparse but this is not something
we can count on.

In contrast, NewsStand CoronaViz deals with variables that are both
time-varying and spatially varying by re-examining the dimensionality
of the data in the sense that a time slider is a natural
representation of one-dimensional data (i.e., time) while a
two-dimensional map is a natural representation of two-dimensional
data.  The problem is how we we represent the values of the variables.
One possible solution is via a histogram but this leads to clutter
on the display and is cumbersome when the data is not spatially
sparse.  Moreover, there may be a layout problem here in the sense
that we cannot allow the histograms to overlap.  An alternative
common solution with the same overlap issues is to use solid
concentric circles where the radii of the circles correspond to the
value and the color corresponds to the identity of the variable.
This type of visualization is known in cartography as a proportional
symbol map~\cite{slocum_thematic_2009}. The 
problem here is when we have multiple dynamic variables as
is the case for our application, then only the one with the largest magnitude can be viewed.
One solution is to vary the colors of the circles but if this method
is used, then we must pay close attention to the order in which we
display the circles so that the one with the largest radius is
displayed first and the remaining circles be displayed in decreasing
order of radius values (this is analogous to the ``back-to-front''
z-buffer display algorithm used in computer graphics).
We can avoid the need to worry about the order in which we display
the circles by using hollow concentric circles where again the color
indicates the identity of the variable while the radius corresponds to
a scaled variable magnitude value.  We use the term {\em geo-circle}
to describe this approach.

The visual strain posed by having a large number of circles can be
relieved by drawing the circles using broken lines of the same width.
At times, the width of the broken lines can be increased with the goal
of drawing attention to a particular set of concentric circles (i.e., a
location whose variable values at a particular instance of time) which
is of interest. We do this in the case of a hover operation wile
panning on the map to show the spatially closest location with nonzero
variable values.  This operation is common in computer graphics where
it is known as a ``pick'' operation (e.g.,
see~\cite{Fole90}).

NewsStand CoronaViz makes use of 6 dynamic variables comparing the
number of confirmed cases. Active cases, recoveries (although not
reported by all jurisdictions), and deaths, and also the
number of news articles and tweets that mention Coronavirus.
Concentric circles (i.e., geo-circles) are used for 4 disease related
variable values while concentric solid circles with different colors,
depending on the number of documents and the nature of the document.
It is advised to just display one of the news article or tweet
variables in which the count is displayed in the circle.  Otherwise
no count is displayed.

The concentric circles make is easy to spot trends and similar values
on the map by looking at the magnitude of the radii.  Other observations
of interest involve trends such as noting lower confirmed case and
death counts over time as the circles get smaller.  Another
encouraging trend is when confirmed case counts become smaller than death counts,
In essence, we are speaking about when concentric circles intersect
and change their relative order.  Of course this must be treated
with caution as the magnitudes of the variables change).

There are a number of ways of presenting the variable values.  The
default in our case is of a cumulative nature.  However, it is
possible to normalize the values over population, or even area.
Normalizing over the area is of possible interest as it could be
used to see if densely populated areas are more likely to lead to
higher incidences of COVID-19 and deaths.

Our goal is to endow NewsStand CoronaViz with a full compliment
of queries that are consistent with its role as a spatio-temporal
and spatio-textual database.  First of all, we have two types of
queries:
\begin{enumerate}
\tightlist
\item location-based:  given a location, what features are present
as well as what are the values of certain variables.  In the
case of documents, we are looking for documents that mention
a particular location name.
\item feature-based:  given a feature, where is it located.
This is also known as spatial data mining.
For example, given a tweet or a news article, what locations
does it mention.  In NewsStand CoronaViz we might be looking for
locations where there are no deaths.
\end{enumerate}

The location-based queries are supported by the ability to pan the
map with a hover operation and always returning the variable values
with the nearest location for which we have data.
The feature-based queries are supported by the NewsStand and
TwitterStand databases that support the news and tweet 
document system.  Feature-based queries require the use of a
pyramid-like data structure on each of the disease-related dynamic
queries.

The animation window is a very important feature as it enables
the execution of a range query where the range is temporal.  Users
can vary the start and end times as well as the animation step size.
In addition users can specify what statistic is being computed
for the window.  It can be cumulative, a time instance like a daily,
weekly, monthly, or any period of days.  Average values for the
window can also be computed.  This is particularly useful for the
``reopen'' discussion which is often based on a rolling 
weekly daily average computation.

Spatial range (also known as window) queries are also of great
interest.  In this case, users use pan and zoom operations to
get a map that is focused on a particular spatial region (e.g.,
the minimum bounding areal box that contains Italy.
Not that in this case, there is also overlap with 
San Marino and the Holy See (i.e., the Vatican in Rome).  In this case
we display the values of the dynamic variables for all
three of these entities.  To get just the values for Italy, users
must zoom in further so that San Marino and the Holy See are not
in the window.  Alternatively if users only want Italy, then they
could simply pose the textual query for which an appropriate
index exists.  Note that as NewsStand CoronaViz zooms into a
region, it has access to more data (as low as county level data).
In the case of tweets and news articles, the area spanned
by the location for which we have this data is not necessarily
coincident with the location for which we have disease-related data
but they are close and thus we report the variable's values
over all the regions that overlap the region of the document related
location for which we have no data (e.g., Egypt and Libya for Benghazi).

Notice that NewsStand CoronaViz enables the execution of the full
compliment of spatio-temporal queries as it supports keeping location
fixed while varying time via the time slider, keeping time fixed and
letting location vary via the hover, panning, and zooming operations.
We can also pick any range of time or space.  The full compliment of 
spatio-textual queries is possible as users can simply go to a
location on the map and obtain the relevant documents, while also
being able to ask for all documents mentioning a particular
keyword.  They can also take advantage of spatial synonyms
when they don't know the exact name of the location of interest.
For example, when seeking a ``Rock Concert in Manhattan,'' concerts in
Harlem, New York City, and Brooklyn are all good answers because
of being contained in Manhattan, containing Manhattan, and being a
spatially adjacent borough, respectively.  This is an example
of a proximity query which we saw previously via the use of a hover
operation in the case of spatial proximity, and the time slider 
in the case of temporal proximity.  Note that in the case of 
temporal proximity, we provide the capability to halt an animation at
arbitrary time instances as well as resuming or terminating it.
In addition, users are also able to set the speed of the animation,
as well as to step through an animation by a specific time interval both
forward and backward in time.

\hypertarget{relatedwork}{%
\section{Related Work}\label{relatedwork}}

In this section we first briefly consider prior work dealing with the
visualization  spatiotemporal data and then review a number of existing systems
designed specifically for monitoring the spread of COVID-19.

\hypertarget{spatiotemporal}{%
\subsection{Spatiotemporal Data Visualization}\label{spatiotemporal}}

Visualization and analysis of temporally varying geospatial data is a difficult
task; as such, it has been the subject of substantial prior work. The
difficulty comes from the inherently multidimensional nature of the data: there
are at a minimum two spatial dimensions and one temporal dimension, in addition
to the dimensionality added by the actual variables being visualized. All of
these dimensions must be projected onto a two dimensions screen. We can broadly
break spatiotemporal visualization techniques into two groups: those that use
animation to capture the time dimensions, and those that attempt to encode
temporally varying information into a single static visualization.

An example of this second variant is presented by \citet{du_banded_2018} who
modify the traditional choropleth map to encode temporal information inside
each area unit. Rather than picking a single color for each areal unit, units
are divided either into bands of either equal width or equal area. Each band
is then assigned a color in the same way areal units are assigned colors in
traditional choropleth maps (e.g.~\citet{slocum_thematic_2009}).

\citet{li_cope_2019} do not use a fully animated approach, but neither do they
commit to showing the full temporal data range in a single image. Instead, they
use an interface termed the ``Event View'' to display images generated for
discrete time intervals side-by-side. To link these images together into a
single cohesive visualization, the authors overlay a ``trend line'' that
connects the time intervals. This trend line is used to link events extracted by
a separate component of their system.

Very often a temporal variant of a well known cartographic visualization
technique can be obtained by applying the existing technique to data within a
time window for series of time window. An animation is obtained by collecting
the individual visualization and displaying them in order by time. This is
approach the basis of \citet{ouyang_algorithms_2000} who develop an algorithm
to generate spatiotemporal cartogram animations. 

\hypertarget{existing systems}{%
\subsection{Existing COVID-19 Monitoring Systems}\label{existing systems}}

In this subsection we compare our visualization tool with some existing systems
recently developed for monitoring the spread of COVID-19. These systems are
described below with an emphasis on pointing out their drawbacks.  

\begin{enumerate}
\tightlist
\item \url{https://coronavirus.jhu.edu/} Coronavirus COVID-19 global cases (Johns Hopkins)
\item \url{https://www.healthmap.org/ncov2019/} Novel coronavirus (COVID-19) outbreak timeline map (HealthMap)
\item \url{https://news.google.com/covid19/map} (Google News)
\item \url{https://hgis.uw.edu/virus/} Novel coronavirus infection map (University of Washington)
\item \url{http://nssac.bii.virginia.edu/covid-19/dashboard/} COVID-19 surveillance dashboard (University of Virginia)
\item \url{https://covid19.who.int/} Novel coronavirus (COVID-19) situation dashboard (WHO)
\item \url{https://www.cdc.gov/coronavirus/2019-ncov/cases-in-us.html} Coronavirus disease 2019 (COVID-19) in the US (CDC)
\item \url{https://www.ecdc.europa.eu/en/geographical-distribution-2019-ncov-cases} Geographical distribution of COVID-19 cases worldwide (ECDC)
\item \url{https://www.kff.org/global-health-policy/fact-sheet/coronavirus-tracker/} COVID-19 coronavirus tracker (Kaiser Family Foundation)
\item \url{https://www.worldometers.info/coronavirus/} COVID-19 coronavirus outbreak (Worldometer)
\item \url{https://multimedia.scmp.com/infographics/news/china/article/3047038/wuhan-virus/index.html} Coronavirus: the new disease Covid-19 explained (South China Morning Post)
\item \url{https://storymaps.arcgis.com/stories/4fdc0d03d3a34aa485de1fb0d2650ee0} Mapping the Wuhan coronavirus outbreak (Esri StoryMaps)
\item \url{https://public.flourish.studio/visualisation/1539110} (Flourish)
\item \url{https://coronavirus.1point3acres.com/en} (1point3acres)
\item \url{https://geods.geography.wisc.edu/covid19/physical-distancing/} (University of Wisconsin)
\end{enumerate}

The Johns Hopkins system 
tabulates cumulative numbers of confirmed, active, deaths, and 
recoveries.  The cumulative numbers of confirmed and active
cases in some of the countries are displayed on the map for some of
the larger countries (in terms of area).  A drawback of the maps is that
zooming in on the map simply increases the resolution of the map but
does not show the data for additional countries.  This is a common
drawback of many of the systems that have been created for visualizing
the coronavirus.  In contrast, the NewsStand approach yields more
article clusters as you zoom in.  The associated article clusters with
the zoomed in area are not as important/relevant as the article
clusters associated with the zoomed out area. 

The HealthMap system
shows the spread of the disease by tabulating the number of new
confirmed cases of the disease on a daily basis and displaying 
it with a circle of a particular size and color anchored
at the location where it was reported (e.g.,~a city,
state, country, etc.). HealthMap still has the drawback that
zooming in only increases the resolution of the map but does not
show a finer allocation of the tabulated properly to the location.

The Google News system makes use of a map query interface and allows
zooming in and reports the variable values for the smaller subunits.
It uses a hover operation to yield the variable values for the
spatial unit being hovered over, as well as disease-related news at times.
It does not have the ability to provide variable values for a
combination of units that make up the viewing window when 
these units are small (e.g., counties) or bigger (countries) as is
done in CoronaViz.  It is static as it has no temporal component other
than precomputed graphs of variable values over a predetermined range
of days unlike CoronaViz where the range is set by the user.

The University of Washington system shows the total number of
confirmed cases, deaths, and recovered for the countries of the world as one
pans the world map.  For the US, zooming in has a greater granularity 
and results in showing how the number of confirmed cases are spatially
distributed in each state.  Descriptive data is also provided 
for the confirmed individuals when the region is sufficiently small.

The Flourish system enables the visualization of just one dynamic
variable such as the number of confirmed cases in a number of countries
at the same instance of time.  Although the data is
spatially-referenced by name (i.e., the names of the countries) no use
is made of a map nor are there any input or output controls.  The
one advantaged of the system is that it is fast which conveys the
urgency of the need to stop its spread.

Both the 1point3acres and Worldometer systems provide comprehensive data and
graphs for the dynamic variables but no animation or maps.  The
dynamic aspect of the variables is captured by the various plots of
the variable values and combinations thereof.  They make a distinction
between cumulative variable values as well as new values.  The 1point3acres
system prides itself in its data collection ability and is more focused
on the virus while the Worldometer system also provides statistics
related to the impact of the disease such as unemployment.

The University of Virginia system displays the number of cumulative
confirmed cases, deaths, and recovered over time using a time slider.  The
countries are colored according to the range of the number of
individuals for the variable being displayed.  Zooming in results in
more locations being placed on the map as well as the inconsistent
decomposition into smaller units such as states for the US and
provinces for China but not for Canada or Australia.

The remaining systems are quite similar in that they only
map the number of confirmed cases in each country in the case of the
WHO and ECDC systems while and in each state for the CDC system.
The Kaiser Family Foundation system also maps the deaths.
None of the WHO, ECDC, CDC, and the Kaiser Family Foundation systems
permit zooming in to get additional data.  Non-interactive maps are
used to tell the story of the coronavirus outbreak in the South China
Post using ESRI StoryMaps.  Instead of the disease-related variables
some systems like that from the University of Wisconsin looks at
a variable that monitors the mobility of the population with
a map query interface that makes use of cell phone data.

\hypertarget{data-collection}{
\section{Data Collection}\label{data-collection}}

Our application contains three distinct types of data: new article data,
twitter data, and official case data. New article and twitter data is collected
as described in the following subsections. Official case data is aggregated by
\citet{dong2020interactive} who ultimately source it from government
organizations who publicly release this information.

\hypertarget{geocoding-news-keywords}{%
\subsection{News Articles}\label{geocoding-news-keywords}}

Two steps are required to obtain geocoded news keywords: important
keywords must be extracted from news articles, and the news articles
must be geocoded by identifying and resolving toponyms to specific
geographic coordinates. Once we have identified a set of keywords and a
set of geographic locations in an article, we take the cross product of
these sets as the set of geocoded keywords. In other words, we create a
pair associating every keyword in the article with every location
mentioned in the article. Each of the keyword location pairs is also
associated with the time of publication of the article in order to
enable the temporal component of our application.

\hypertarget{keyword-extraction}{%
\subsubsection{Keyword Extraction}\label{keyword-extraction}}

The extraction of news keywords is handled primarily by the original NewsStand
system and several prior extensions to the system ~\cite{Abde15, Lan12, Lan14,
Waji16}.  A brief overview is given here, but the original publications should
be referred to for a detailed explanation.

The original NewsStand implementation takes the simplest approach to keyword
extraction. Keywords are chosen based on the TD-IDF scores for terms in the
document that are computed for document clustering.  Words with high TF-IDF
scores occur more frequently in an article relative to their frequency in the
entire corpus.  As such, these words are generally important to the content of
an article and are a good approximation of the article's keywords.  The word
with the highest TF-IDF score for a document is therefore the best keyword, and
more keywords can be obtained by selecting words with progressively lower
scores.  While this approach captures the broadest range of keywords, it may be
desirable to limit keywords to a specific domain. Techniques for this have been
implemented by extension to NewsStand.

\citet{Lan12, Lan14} used NewsStand to visualize
the progress of potential outbreaks of a disease as measured by the geographic
extent of document clusters that mention the disease.  While this work
incorporates spatio-temporal analysis into NewsStand, it is restricted to
tracking news related to diseases and it only tracks the growth and geographic
distribution of a single cluster of documents in relation to a disease.  In
contrast, the application developed for this paper is able to track the spread
of arbitrary keywords or topics, including diseases, across all documents in
the NewsStand database, and it incorporates into its visualization data for
case numbers as they are officially available.  This extra capacity makes the
application more widely applicable than prior work.

\citet{Abde15} implemented the ability to detect prominent
mentions of different brands and companies within news articles.  To accomplish
this, the authors created two rule based classifiers and trained one supervised
machine learning classifier.  The goal of all three classifiers was to decide
if a brand is mentioned prominently in an article given the name of the brand
and the local context in which the brand is mentioned. 

\citet{Waji16} developed a classification technique for determining if a news
article discusses criminal activity and extracting keywords from the article
that are directly related to the crime discussed.  To accomplish this the
authors first selectecd articles that contained one of a set of predetermined
keywords that are likely indicative of articles about criminal activity (e.g.
murder or theft) before classifying each of these article as either primarily
about crime or not about crime using a support vector machine (SVM) classifier.
This approach is very similar to that of \citet{Abde15}.

The current implementation of CoronaViz makes use only of the TF-IDF keywords,
as we apply a strict query term filter to these keywords making more involved
preprocessing unweary. The system as implemented can be modified to make use of
any of the existing keyword extraction methods if that is desirable for a
specific application. 

For the purpose of this application, once keywords have been extracted
from article text, we examine only those that we believe will be most
relevant to the disease we are tracking. For instance, we may
specifically look for news articles containing the keyword
``coronavirus''. This approach should achieve high precision, since it
is unlikely that an article unrelated to the virus will contain this
keyword, but it may suffer from low recall.

\hypertarget{geocoding}{%
\subsubsection{Geocoding}\label{geocoding}}

Geocoding is the process of associating concrete geographic information
(i.e.~latitude longitude pairs) with a piece of text. Geocoding can be framed
as a specific variant of the keyword extraction task in the sense that we must
first find keywords that are likely to refer to geographic locations and then
decide which of these locations are important enough to associate with the
documents~\cite{Lieb11}. The third step in geocoding which
is not required for general keyword extraction is toponym resolution. In
toponym resolution, a toponym must be assigned a single latitude longitude pair
to be its final location. This is nontrivial because there many ambiguous
toponyms that can be used to refer to multiple distinct locations (e.g.~Paris,
London, etc.)~\cite{Lieb12}.

\subsection{Twitter}

Alongside news articles, we also collect data from and track term usage in
tweets. Our approach to this is slightly different from what is used for news
articles because tweets are much shorter and rarely contain the implicit
geographic information that is used to geocode news articles. To monitor term
usage, we only check that the query term appears in the tweet at some point
rather than extracting keywords based on TF-IDF scores. To assign geographic
coordinates to the tweets, we depend on tweets that are explicitly geotagged by
the Twitter user.

\hypertarget{application-interface}{%
\section{Application Interface}\label{application-interface}}

\begin{figure*}
  \centering
  \includegraphics[width=.9\textwidth]{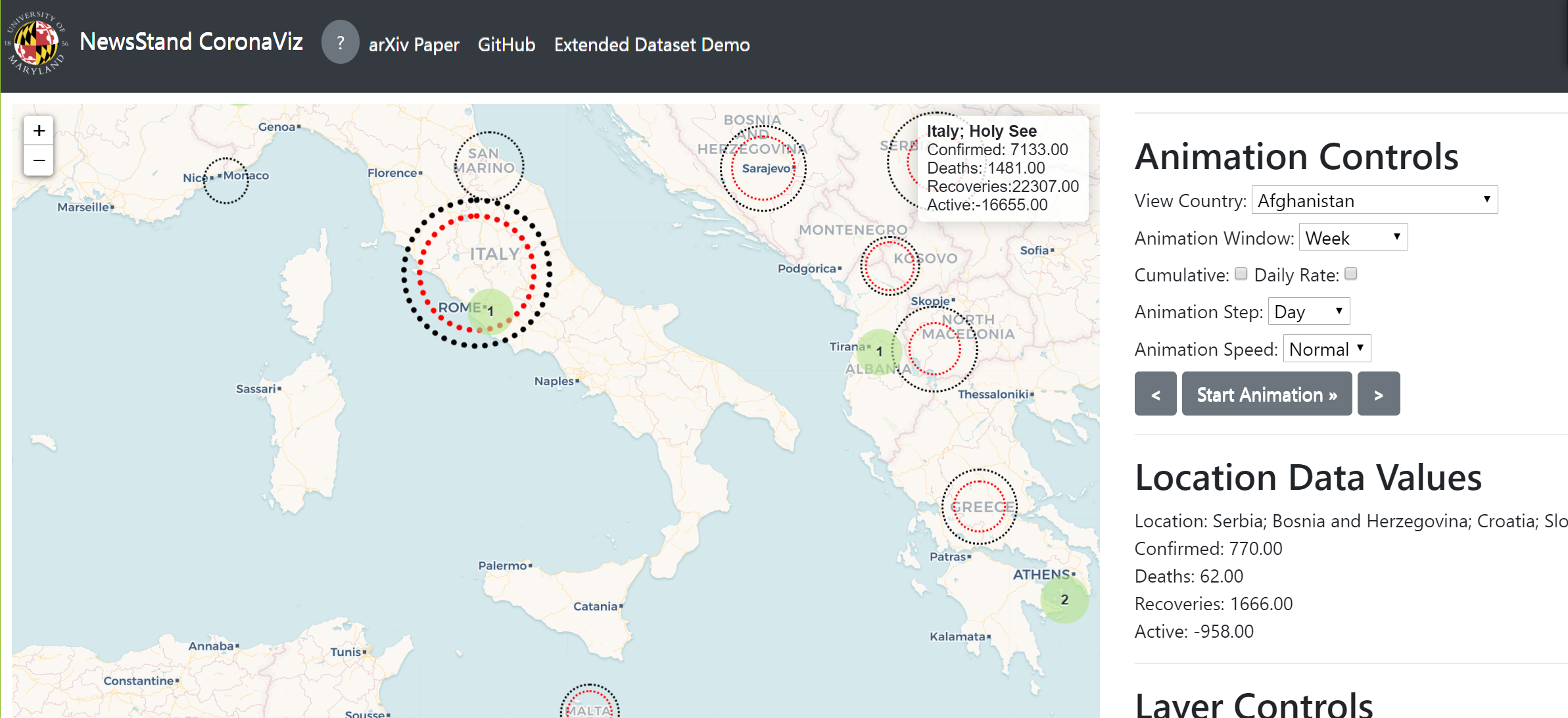}
  \caption{Example NewsStand CoronaViz screenshot for Italy and the Balkan States}
  \label{fig:italy}
\end{figure*}

\begin{figure}
  \centering
  \includegraphics[width=.5\columnwidth]{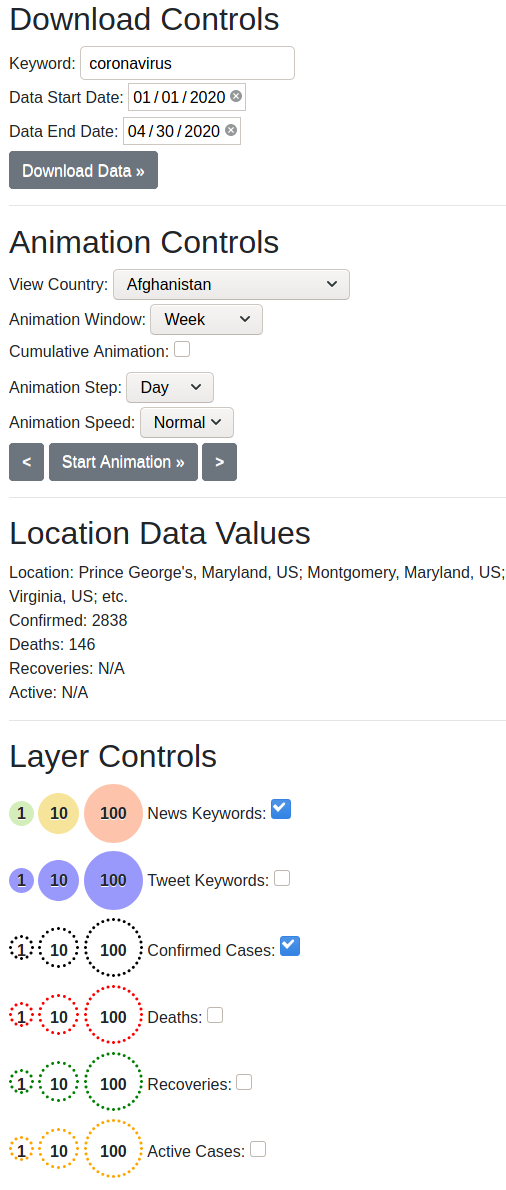}
  \caption{Screenshot showing the full contents of the controls panel.}
  \label{fig:controls}
\end{figure}

Our application is implemented as an interactive website using HTML, CSS, and
JavaScript that communicates with our database through a Ruby On Rails web
server. It consists of an interactive map query interface that is used to
display our data and animations. In addition, a collection of controls are used
to download data from our server and to select how it is displayed on the map.
A screenshot of our application is shown in Figure~\ref{fig:screenshot}, and a
more detailed screenshot showing the full control panel is in
Figure~\ref{fig:controls}. A second screenshot showing CoronaViz used to
visualize data on smaller scale (Italy and the Balkan States) is shown in
Figure~\ref{fig:italy}.

\hypertarget{map-interface}{%
\subsection{Map Query Interface}\label{map-interface}}

Our map query interface is built around an interactive web map provided by the
Leaflet JavaScript Library.\footnote{https://leafletjs.com/} Data is rendered
onto this map using a technique called marker clustering, which is implemented
by an extension to
Leaflet.\footnote{https://github.com/Leaflet/Leaflet.markercluster} Marker
clustering allows large numbers of points to be rendered quickly without
overloading the user with information. Rather than rendering each point
individually, points are clustered into aggregate markers. As the user
increases the zoom level of the map, focusing on a smaller area, these
aggregate markers are split into two or more new markers that together
represent all the points represented by the original marker. This decomposition
allows more detail to be displayed when examining a small area without showing
too much detail at lower zoom levels.

The interface can display five different data layers: news keywords, twitter
keywords, confirmed COVID-19 cases, deaths from COVID-19, and recovered
COVID-19 cases. Data for the news and twitter layers was collected as described
in Section~\ref{data-collection} while data used for the remaining layers was
originally gathered by \citet{dong2020interactive}. Layers can be viewed
individually or simultaneously to visualize the spatial and temporal
relationship between the variables in each layer.

Each marker represents some collection of points, so the size and color of the
markers are scaled to represent the magnitude of this collection. Markers
representing a single point are set to have a radius of \(40\) pixels. From
there, the marker's radius scales with the squared logarithm of the number of
points ( \(r_M = 40 + \log^2 2 \lvert M \rvert\)). Markers for the news data
layer also have their color scaled depending on the number points.
Green is used for markers representing up to ten points, yellow is used for
those that represent up to \(100\) points, and red is used for all larger
markers. These functions are chosen to be visually appealing, but there is no
other concrete reason for their selection.

The map query interface also gives access to the underlying data for each layer.
When the user hovers their mouse over a marker, an overlay on the interface
is updated to show the exact data values for that marker or marker cluster.
When a marker is clicked, this data is also displayed in a in a secondary information box
located directly below the animation controls on the right side of our
application. This placement is intended to facilitate monitoring changes in
data values as the user manipulates the incremental animation controls
described in the next section.

\hypertarget{interface-controls}{%
\subsection{Interface Controls}\label{interface-controls}}

The first control located below the map interface is a slider bar that can be
manipulated to control the range of dates in the data displayed on the map.
Moving either knob on the slider causes the map interface to be updated. The
exact dates selected are displayed below the slider.  Below the date range
slider there are three panels labeled "Download Controls," "Animation
Controls," and "Layer Controls."

The download controls are used to retrieve data from our database.  First,
there is a text entry box labeled ``keyword'' in which the user types a keyword
that they want to visualize. There are then two date entry fields labeled ``Data
Start Date'' and ``Data End Date''. These control the range of dates for data
that is downloaded from our server. Note that this is not necessarily the same as
the range of dates for data that is rendered on the map. Finally, there is a
button labeled ``Download Data'' that, when clicked, sends a request to our
server for data within the specified time range matching the provided keyword. This
data is rendered on the map query interface, replacing any previous data.

The animation controls deal with using downloaded data to generate time series
animations. Most prominently, there is the ``Start Animation'' button that
initiates the animation when clicked. Clicking this button a second time will
pause an ongoing animation such that it can be resumed from the current position later.
Located to the left and right of this button are two arrow buttons that are
used to move incrementally through the animation. Clicking on the left arrow
steps backwards to the previous frame of the animation while right arrow
advances to the next frame. With the incremental controls, the user can examine
the exact data value for each frame of the animation, enabling more precise 
analysis than would be possible only using the continuous animation mode.

The properties of the animation are controlled by the next four inputs. The
first input controls the size of the range of dates displayed in each frame of
the animation. It can be set to hour, day, week, or month. It can also be set
to a custom size through use of the ``Displayed Data Range'' controls discussed
above. The checkbox below this is used to toggle the cumulative animation mode.
Under the default configuration, only an interval defined by the animation
window is displayed at one time. This behavior is desirable for monitoring
disease spread because it captures where the disease is active at any given
time and not every location that it has ever been active. Nonetheless, it may
be interesting to view disease spread in a cumulative manner, so this
functionality is provided.  Next, there is a selection input for the interval
advanced between each frame of the animation. This can be set to day, week, or
month. The final input field is used to set the speed of the animation to
either "slow", "normal", or "fast". These values correspond to a minimum number
of milliseconds waited between frames of the animation: $500$, $100$, and $0$
milliseconds respectively. While these are minimum delays, the actual interval
may be longer due to time required to compute the subsequent frame.

Finally, the layer controls are used to select between the possible data layers
that can be rendered on the map interface. By default, only the news keyword
and confirmed cases data layers are selected.  Selecting any of the other
layers will cause that layer to be rendered in addition to any previously
selected layers.

When the confirmed cases layer is selected, there are markers representing the
total number of confirmed cases in a location rendered on the map. Similarly
there are layers for recoveries and deaths that contain markers representing
these variables. When multiple of these layers are selected simultaneously, all
selected layers are rendered simultaneously as sets of concentric circles. The
concentric circles allow for the comparison of values for different layers in
a single location by examining the relative size of circles centered at that
location while also allowing for comparison between different locations.

\hypertarget{example-usage}{%
\subsection{Example Usage}\label{example-usage}}

This subsection provides a walk through demonstrating how our application
can be used to obtain an animation to show the geographic change in the
discussion of COVID-19 in news articles over time. The instructions here
are applicable to other similar uses of our application.

To begin, navigate to \url{coronaviz.umiacs.io} to access our website. Once
there the ``Keyword'' entry field should be initially populated with
``coronavirus''.  If this is not the case, select the field and type in this
term. The next step is selecting the range of dates to be viewed. A reasonable
range for visualizing Corona Virus might start in December 2019 and proceed
until the current date. To make this selection, click on and select a date for
the ``Data Start Date'' field. ``Data End Date'' should have been initialized
the current date, but, if it is not, a date should also be selected here. If
the default query term and date range have not been changed, the data should
have been retrieved automatically upon loading the webpage; otherwise, the
``Download Data'' button must be clicked after selecting the custom query term
and data range to download this data from our server.  This may take a
noticeable amount of time depending on network speed and the size of the query
result. Once the download is complete, data will be rendered on the map.

The next step is to configure the animations controls and initiate an animation
sequence. The animation options (``Animation Window'' and ``Animation Step'')
are set to week and day respectively by default.  These should be reasonable
values but, at this point, they can be changed. Clicking the ``Start
Animation'' button starts an animation on the map query interface that begins
by displaying downloaded data with the earliest publication date and proceeds
until the final frame that contains the most recent data. Each frame of the
animation shows a period of time defined by ``Animation Window'' and the start
of the frame is advanced by an interval defined by ``Animation Step'' between
each frame. The animation can be terminated early by clicking the button, now
labeled ``Pause Animation'',  a second time.

\hypertarget{evaluation}{%
\section{Evaluation}\label{evaluation}}

In this section we evaluate the effectiveness of tracking keyword usage in news
articles and tweets to better understand the progression of COVID-19. To do
this we search for correlation between the confirmed case numbers collected by
\cite{dong2020interactive} and the cumulative number of news articles or
tweets recognized by our system.  We include geographic information in our
analysis by limiting it to confirmed cases and news articles in a specific
location. We are able to restrict the analysis in this manner because the
dataset of confirmed cases we use already organizes confirmed case numbers by
location, our news article dataset is geocoded as described in
Section~\ref{geocoding}, and the Tweet dataset is derived from geotagged
tweets. To further investigate the utility of our technique, we perform the
comparison at three geographic scales: world, nation, and region. The results
of our evaluations are summarized in Table~\ref{tbl:correlation} and
Figures~\ref{fig:world-data}, \ref{fig:china-data}, and \ref{fig:us-data}.
\footnote{In each plot of article keyword counts, the
daily number of articles collected is zero from February 14 to February 17,
March 9 to March 11, and on March 22. This was caused by problems in our data
collection system, so the data points are removed when computing the
correlation coefficients in Table~\ref{tbl:correlation}.}

\subsection{News Keywords}

\begin{table}
\centering
\begin{tabular}{lll}
\toprule
      & \multicolumn{2}{c}{Correlation coefficient}\\
Query Area & Keyword \em{coronavirus} & No filter\\
\midrule
World & $0.75$ & $-0.15$\\
China & $-0.11$ & $-0.05$\\
Hubei Province & $-0.12$ & $-0.08$\\
United States & $0.37$ & $-0.11$\\
Washington state & $0.25$ & $0.14$\\
\bottomrule
\end{tabular}
\caption{\label{tbl:correlation} Correlation coefficient between cumulative new articles and confirmed cases at each scale.}
\end{table}

\begin{figure}
  \begin{subfigure}{.5\columnwidth}
    \resizebox{\columnwidth}{!}{\input{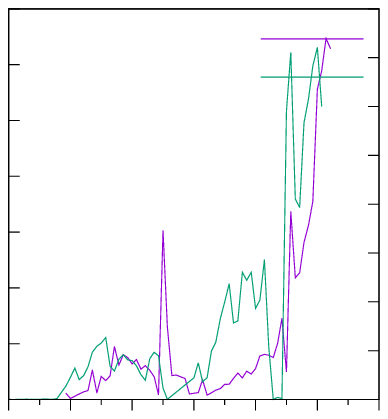}}
    \caption{World filtered}
    \label{fig:world-time-series}
  \end{subfigure}%
  ~
  \begin{subfigure}{.5\columnwidth}
    \resizebox{\columnwidth}{!}{\input{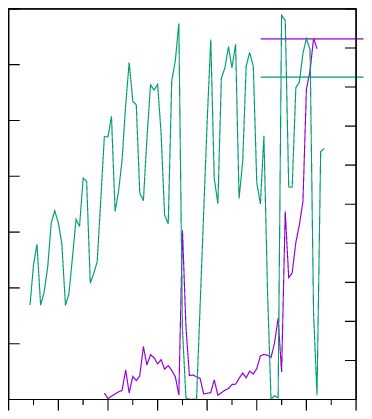}}
    \caption{World unfiltered}
    \label{fig:world-nofilter-time-series}
  \end{subfigure}
  \caption{Filtered and unfiltered data at the global scale.}
  \label{fig:world-data}
\end{figure}

We first ignore specific geographic information extracted from articles and
perform this analysis on a global scale. This does not utilize the geographic
information we collect, but it is interesting as a point of comparison to other
scales. Figure~\ref{fig:world-time-series} plots the daily number of articles
using the keyword ``coronavirus'' as a time series between January 22, 2020 and
March 20, 2020. To demonstrate the utility of news articles in tracking the
progress of the virus, this figure also shows a time series for the daily
number of new confirmed cases world wide.  It is clear that both datasets trend
upwards at similar rates. To determine if correlation exists between article
counts and case numbers, we calculated the Pearson correlation coefficient
between the datasets. We found the coefficient to be $0.75$, indicating that the
datasets are linearly correlated. To investigate the effect of our keyword
filter, we performed a similar comparison where we do not filter news
articles for any specific terms. Figure~\ref{fig:world-nofilter-time-series}
shows the time series plot for this data. This time, the article counts clearly
do not correlate with case numbers. This is realized in a $-0.15$ correlation
coefficient. Since the magnitude of the correlation coefficient is
significantly larger when applying the keyword filter, we can see that it is
useful in this application.

\begin{figure}
  \begin{subfigure}{.5\columnwidth}
    \resizebox{\columnwidth}{!}{\input{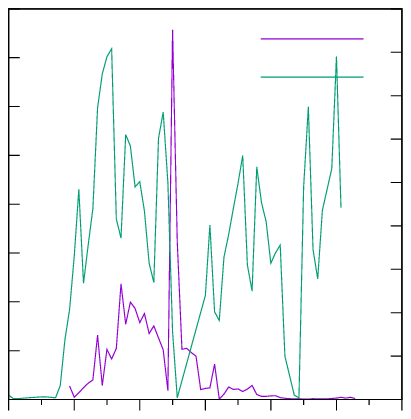}}
    \caption{China filtered}
    \label{fig:cn-time-series}
  \end{subfigure}%
  ~
  \begin{subfigure}{.5\columnwidth}
    \resizebox{\columnwidth}{!}{\input{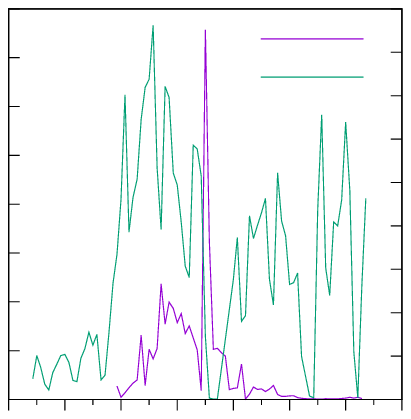}}
    \caption{China unfiltered}
    \label{fig:cn-nofilter-time-series}
  \end{subfigure}

  \centering
  \begin{subfigure}{.5\columnwidth}
    \resizebox{\columnwidth}{!}{\input{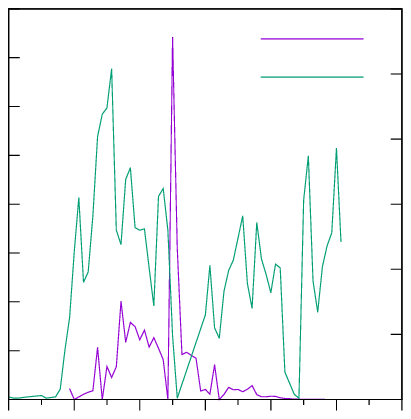}}
    \caption{Hubei filtered}
    \label{fig:hubei-time-series}
  \end{subfigure}%
  ~
  \centering
  \begin{subfigure}{.5\columnwidth}
    \resizebox{\columnwidth}{!}{\input{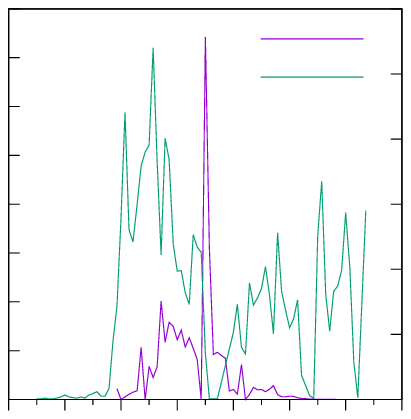}}
    \caption{Hubei unfiltered}
    \label{fig:hubei-nofilter-time-series}
  \end{subfigure}

  \caption{Filtered and unfiltered data in all of China and Hubei province.}
  \label{fig:china-data}
\end{figure}

The plots in Figure~\ref{fig:china-data} are formatted the same as Figure~\ref{fig:world-data}, 
but they focus on smaller geographic scales. In Figure~\ref{fig:cn-time-series},
the plot is restricted to articles geocoded to China and confirmed cases in
China. The correlation coefficient between these datasets is $-0.11$,
indicating that there is no correlation after applying these geographic
filters. Figure~\ref{fig:hubei-time-series} further restricts the data to the
Hubei province of China. The correlation coefficient at this scale is $-0.12$,
similar to found for all of China.  Figures~\ref{fig:cn-nofilter-time-series},
and \ref{fig:hubei-nofilter-time-series} show the plots for China and Hubei
when the keyword filter is not applied. Since there was no correlation found
with the filter applied we do not expect anything to change when it is removed.
As expected, the new correlation coefficients are almost unchanged with
$-0.05$ and $-0.08$ for China and Hubei respectively.

\begin{figure}
  \begin{subfigure}{.5\columnwidth}
    \resizebox{\columnwidth}{!}{\input{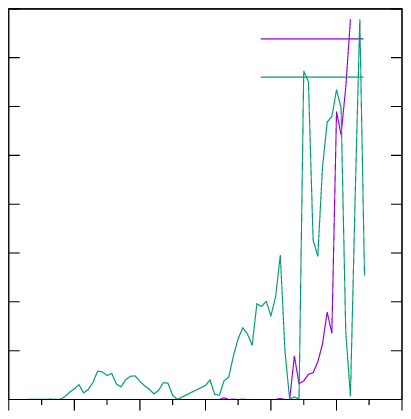}}
    \caption{United States filtered}
    \label{fig:us-time-series}
  \end{subfigure}%
  ~
  \begin{subfigure}{.5\columnwidth}
    \resizebox{\columnwidth}{!}{\input{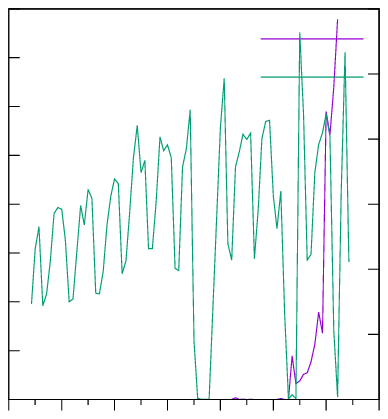}}
    \caption{United States unfiltered}
    \label{fig:us-nofilter-time-series}
  \end{subfigure}

  \centering
  \begin{subfigure}{.5\columnwidth}
    \resizebox{\columnwidth}{!}{\input{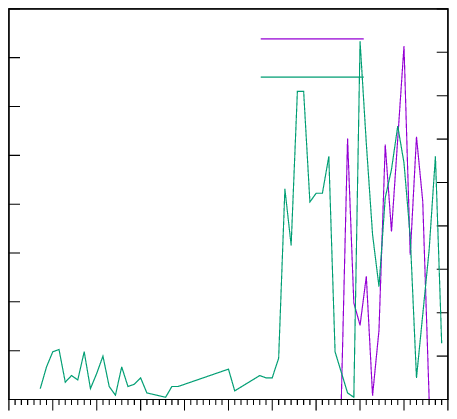}}
    \caption{Washington filtered}
    \label{fig:wa-time-series}
  \end{subfigure}%
  ~
  \begin{subfigure}{.5\columnwidth}
    \resizebox{\columnwidth}{!}{\input{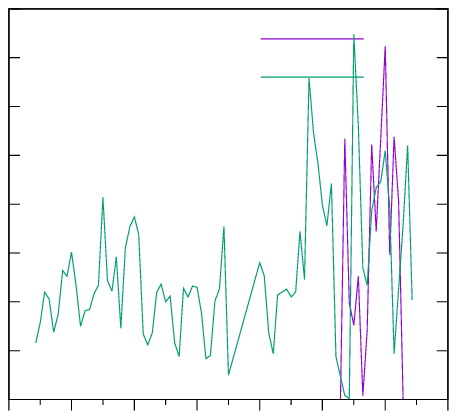}}
    \caption{Washington unfiltered}
    \label{fig:washington-nofilter-time-series}
  \end{subfigure}

  \caption{Filtered and unfiltered data in all of the United States and Washington state.}
  \label{fig:us-data}
\end{figure}

In our final set of evaluations, we shift geographic focus from China to the
United States. Figure~\ref{fig:us-data} contains plots for a country level of China
analysis of the United States and a more specific analysis of Washington state.
In contrast to the country level analysis where no correlation was found, the
correlation coefficient found for the United States ($0.37$) is greater than
that found in the global analysis. While the correlation coefficient calculated
for the Washington data is much lower ($0.25$), it is notably higher than the
values computed for either China or Hubei. The correlation coefficient for the
United States and Washington when not applying the keyword filter ($-0.11$ and
$0.14$ respectively) is comparable to that for of the three previous datasets.

In our evaluation of the news keyword dataset, we found that the number of news
articles using the term "coronavirus" is correlated with the number of new
cases of COVID-19 when correlation is computed at a global scale. When
restricting the analysis to a country wide scale, we found weak correlation
when using data for the United States, but we were not able to find any
correlation in our dataset for China. There is a similar but less pronounced
relationship at the region level. While the Washington state dataset does not
display particularly strong correlation with the confirmed cases dataset, it is
notably higher than the correlation we found for Hubei. A possible explanation
for this inconsistency is that many articles discussing COVID-19 will mention
China at some point since it is the origin of the virus. To overcome this issue
we may be able to develop a technique to determine which term, toponym pairs in
a news article should be included in our dataset. Our dataset currently
contains an entry for every toponym in an article if the article also contains
the query term.

%
%

\hypertarget{conclusion}{%
\section{Concluding Remarks and Future Work}\label{conclusion}}

In this paper we have described the NewsStand CoronaViz system for visualizing 
spatiotemporal data and its specific application to tracking
the progress of the COVID-19 pandemic.

The primary limitation of our current work is in our process for obtaining
spatially referenced mentions of COVID-19 from news articles. Our current
procedure looks for articles that contain the term "coronavirus" as a keyword,
and then finds all toponyms that are mentioned in the article. While this is
good enough to be useful, as demonstrated by the moderate correlation we found,
there is room for improvement. In particular, we plan to investigate more
sophisticated natural language processing techniques for determining if a given
articles discuss the virus and for determining if a specific toponym used in
such an article should be counted as a location mentioned in reference to the
disease. 

We also acknowledge that we have not yet conducted any user study to
objectively determine the effectivness of the system. We plan to design and
conduct a study of our dynamic concentric proportional symbols and the ability of
users to extract information from our system in its animated and incremental
modes.

\bibliographystyle{abbrvnat}
\fontsize{8}{9.5}\selectfont
\setlength{\bibsep}{0pt plus 0.3ex}
\bibliography{corona_tracking,hjs}

\end{document}

%% file: plots/world_filter.tex
\begingroup
  \makeatletter
  \providecommand\color[2][]{%
    \GenericError{(gnuplot) \space\space\space\@spaces}{%
      Package color not loaded in conjunction with
      terminal option `colourtext'%
    }{See the gnuplot documentation for explanation.%
    }{Either use 'blacktext' in gnuplot or load the package
      color.sty in LaTeX.}%
    \renewcommand\color[2][]{}%
  }%
  \providecommand\includegraphics[2][]{%
    \GenericError{(gnuplot) \space\space\space\@spaces}{%
      Package graphicx or graphics not loaded%
    }{See the gnuplot documentation for explanation.%
    }{The gnuplot epslatex terminal needs graphicx.sty or graphics.sty.}%
    \renewcommand\includegraphics[2][]{}%
  }%
  \providecommand\rotatebox[2]{#2}%
  \@ifundefined{ifGPcolor}{%
    \newif\ifGPcolor
    \GPcolortrue
  }{}%
  \@ifundefined{ifGPblacktext}{%
    \newif\ifGPblacktext
    \GPblacktextfalse
  }{}%
  \let\gplgaddtomacro\g@addto@macro
  \gdef\gplbacktext{}%
  \gdef\gplfronttext{}%
  \makeatother
  \ifGPblacktext
    \def\colorrgb#1{}%
    \def\colorgray#1{}%
  \else
    \ifGPcolor
      \def\colorrgb#1{\color[rgb]{#1}}%
      \def\colorgray#1{\color[gray]{#1}}%
      \expandafter\def\csname LTw\endcsname{\color{white}}%
      \expandafter\def\csname LTb\endcsname{\color{black}}%
      \expandafter\def\csname LTa\endcsname{\color{black}}%
      \expandafter\def\csname LT0\endcsname{\color[rgb]{1,0,0}}%
      \expandafter\def\csname LT1\endcsname{\color[rgb]{0,1,0}}%
      \expandafter\def\csname LT2\endcsname{\color[rgb]{0,0,1}}%
      \expandafter\def\csname LT3\endcsname{\color[rgb]{1,0,1}}%
      \expandafter\def\csname LT4\endcsname{\color[rgb]{0,1,1}}%
      \expandafter\def\csname LT5\endcsname{\color[rgb]{1,1,0}}%
      \expandafter\def\csname LT6\endcsname{\color[rgb]{0,0,0}}%
      \expandafter\def\csname LT7\endcsname{\color[rgb]{1,0.3,0}}%
      \expandafter\def\csname LT8\endcsname{\color[rgb]{0.5,0.5,0.5}}%
    \else
      \def\colorrgb#1{\color{black}}%
      \def\colorgray#1{\color[gray]{#1}}%
      \expandafter\def\csname LTw\endcsname{\color{white}}%
      \expandafter\def\csname LTb\endcsname{\color{black}}%
      \expandafter\def\csname LTa\endcsname{\color{black}}%
      \expandafter\def\csname LT0\endcsname{\color{black}}%
      \expandafter\def\csname LT1\endcsname{\color{black}}%
      \expandafter\def\csname LT2\endcsname{\color{black}}%
      \expandafter\def\csname LT3\endcsname{\color{black}}%
      \expandafter\def\csname LT4\endcsname{\color{black}}%
      \expandafter\def\csname LT5\endcsname{\color{black}}%
      \expandafter\def\csname LT6\endcsname{\color{black}}%
      \expandafter\def\csname LT7\endcsname{\color{black}}%
      \expandafter\def\csname LT8\endcsname{\color{black}}%
    \fi
  \fi
    \setlength{\unitlength}{0.0500bp}%
    \ifx\gptboxheight\undefined%
      \newlength{\gptboxheight}%
      \newlength{\gptboxwidth}%
      \newsavebox{\gptboxtext}%
    \fi%
    \setlength{\fboxrule}{0.5pt}%
    \setlength{\fboxsep}{1pt}%
\begin{picture}(4752.00,3484.00)%
    \gplgaddtomacro\gplbacktext{%
      \csname LTb\endcsname
      \put(1078,1013){\makebox(0,0)[r]{\strut{}$0$}}%
      \put(1078,1334){\makebox(0,0)[r]{\strut{}$5000$}}%
      \put(1078,1656){\makebox(0,0)[r]{\strut{}$10000$}}%
      \put(1078,1977){\makebox(0,0)[r]{\strut{}$15000$}}%
      \put(1078,2299){\makebox(0,0)[r]{\strut{}$20000$}}%
      \put(1078,2620){\makebox(0,0)[r]{\strut{}$25000$}}%
      \put(1078,2942){\makebox(0,0)[r]{\strut{}$30000$}}%
      \put(1078,3263){\makebox(0,0)[r]{\strut{}$35000$}}%
      \put(1210,818){\rotatebox{-45}{\makebox(0,0)[l]{\strut{}01/09}}}%
      \put(1566,818){\rotatebox{-45}{\makebox(0,0)[l]{\strut{}01/23}}}%
      \put(1921,818){\rotatebox{-45}{\makebox(0,0)[l]{\strut{}02/06}}}%
      \put(2277,818){\rotatebox{-45}{\makebox(0,0)[l]{\strut{}02/20}}}%
      \put(2632,818){\rotatebox{-45}{\makebox(0,0)[l]{\strut{}03/05}}}%
      \put(2988,818){\rotatebox{-45}{\makebox(0,0)[l]{\strut{}03/19}}}%
      \put(3343,818){\rotatebox{-45}{\makebox(0,0)[l]{\strut{}04/02}}}%
      \put(3475,1013){\makebox(0,0)[l]{\strut{}$0$}}%
      \put(3475,1294){\makebox(0,0)[l]{\strut{}$1000$}}%
      \put(3475,1576){\makebox(0,0)[l]{\strut{}$2000$}}%
      \put(3475,1857){\makebox(0,0)[l]{\strut{}$3000$}}%
      \put(3475,2138){\makebox(0,0)[l]{\strut{}$4000$}}%
      \put(3475,2419){\makebox(0,0)[l]{\strut{}$5000$}}%
      \put(3475,2701){\makebox(0,0)[l]{\strut{}$6000$}}%
      \put(3475,2982){\makebox(0,0)[l]{\strut{}$7000$}}%
      \put(3475,3263){\makebox(0,0)[l]{\strut{}$8000$}}%
    }%
    \gplgaddtomacro\gplfronttext{%
      \csname LTb\endcsname
      \put(209,2138){\rotatebox{-270}{\makebox(0,0){\strut{}Confirmed Cases}}}%
      \put(4245,2138){\rotatebox{-270}{\makebox(0,0){\strut{}Article Count}}}%
      \put(2276,154){\makebox(0,0){\strut{}Day and Month (2020)}}%
      \csname LTb\endcsname
      \put(2530,3090){\makebox(0,0)[r]{\strut{}Confirmed}}%
      \csname LTb\endcsname
      \put(2530,2870){\makebox(0,0)[r]{\strut{}Article}}%
    }%
    \gplbacktext
    \put(0,0){\includegraphics{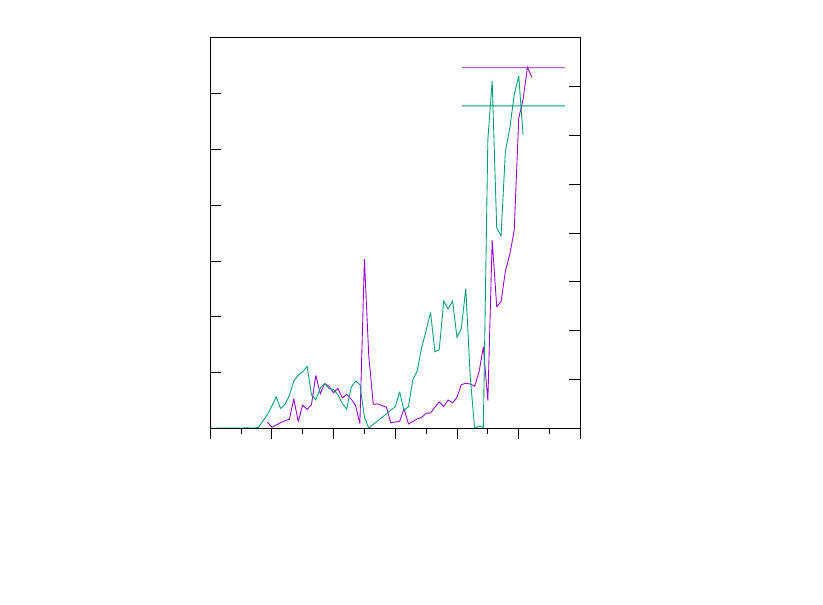}}%
    \gplfronttext
  \end{picture}%
\endgroup

%% file: plots/world_nofilter.tex
\begingroup
  \makeatletter
  \providecommand\color[2][]{%
    \GenericError{(gnuplot) \space\space\space\@spaces}{%
      Package color not loaded in conjunction with
      terminal option `colourtext'%
    }{See the gnuplot documentation for explanation.%
    }{Either use 'blacktext' in gnuplot or load the package
      color.sty in LaTeX.}%
    \renewcommand\color[2][]{}%
  }%
  \providecommand\includegraphics[2][]{%
    \GenericError{(gnuplot) \space\space\space\@spaces}{%
      Package graphicx or graphics not loaded%
    }{See the gnuplot documentation for explanation.%
    }{The gnuplot epslatex terminal needs graphicx.sty or graphics.sty.}%
    \renewcommand\includegraphics[2][]{}%
  }%
  \providecommand\rotatebox[2]{#2}%
  \@ifundefined{ifGPcolor}{%
    \newif\ifGPcolor
    \GPcolortrue
  }{}%
  \@ifundefined{ifGPblacktext}{%
    \newif\ifGPblacktext
    \GPblacktextfalse
  }{}%
  \let\gplgaddtomacro\g@addto@macro
  \gdef\gplbacktext{}%
  \gdef\gplfronttext{}%
  \makeatother
  \ifGPblacktext
    \def\colorrgb#1{}%
    \def\colorgray#1{}%
  \else
    \ifGPcolor
      \def\colorrgb#1{\color[rgb]{#1}}%
      \def\colorgray#1{\color[gray]{#1}}%
      \expandafter\def\csname LTw\endcsname{\color{white}}%
      \expandafter\def\csname LTb\endcsname{\color{black}}%
      \expandafter\def\csname LTa\endcsname{\color{black}}%
      \expandafter\def\csname LT0\endcsname{\color[rgb]{1,0,0}}%
      \expandafter\def\csname LT1\endcsname{\color[rgb]{0,1,0}}%
      \expandafter\def\csname LT2\endcsname{\color[rgb]{0,0,1}}%
      \expandafter\def\csname LT3\endcsname{\color[rgb]{1,0,1}}%
      \expandafter\def\csname LT4\endcsname{\color[rgb]{0,1,1}}%
      \expandafter\def\csname LT5\endcsname{\color[rgb]{1,1,0}}%
      \expandafter\def\csname LT6\endcsname{\color[rgb]{0,0,0}}%
      \expandafter\def\csname LT7\endcsname{\color[rgb]{1,0.3,0}}%
      \expandafter\def\csname LT8\endcsname{\color[rgb]{0.5,0.5,0.5}}%
    \else
      \def\colorrgb#1{\color{black}}%
      \def\colorgray#1{\color[gray]{#1}}%
      \expandafter\def\csname LTw\endcsname{\color{white}}%
      \expandafter\def\csname LTb\endcsname{\color{black}}%
      \expandafter\def\csname LTa\endcsname{\color{black}}%
      \expandafter\def\csname LT0\endcsname{\color{black}}%
      \expandafter\def\csname LT1\endcsname{\color{black}}%
      \expandafter\def\csname LT2\endcsname{\color{black}}%
      \expandafter\def\csname LT3\endcsname{\color{black}}%
      \expandafter\def\csname LT4\endcsname{\color{black}}%
      \expandafter\def\csname LT5\endcsname{\color{black}}%
      \expandafter\def\csname LT6\endcsname{\color{black}}%
      \expandafter\def\csname LT7\endcsname{\color{black}}%
      \expandafter\def\csname LT8\endcsname{\color{black}}%
    \fi
  \fi
    \setlength{\unitlength}{0.0500bp}%
    \ifx\gptboxheight\undefined%
      \newlength{\gptboxheight}%
      \newlength{\gptboxwidth}%
      \newsavebox{\gptboxtext}%
    \fi%
    \setlength{\fboxrule}{0.5pt}%
    \setlength{\fboxsep}{1pt}%
\begin{picture}(4752.00,3484.00)%
    \gplgaddtomacro\gplbacktext{%
      \csname LTb\endcsname
      \put(1078,1013){\makebox(0,0)[r]{\strut{}$0$}}%
      \put(1078,1334){\makebox(0,0)[r]{\strut{}$5000$}}%
      \put(1078,1656){\makebox(0,0)[r]{\strut{}$10000$}}%
      \put(1078,1977){\makebox(0,0)[r]{\strut{}$15000$}}%
      \put(1078,2299){\makebox(0,0)[r]{\strut{}$20000$}}%
      \put(1078,2620){\makebox(0,0)[r]{\strut{}$25000$}}%
      \put(1078,2942){\makebox(0,0)[r]{\strut{}$30000$}}%
      \put(1078,3263){\makebox(0,0)[r]{\strut{}$35000$}}%
      \put(1210,818){\rotatebox{-45}{\makebox(0,0)[l]{\strut{}12/26}}}%
      \put(1496,818){\rotatebox{-45}{\makebox(0,0)[l]{\strut{}01/09}}}%
      \put(1782,818){\rotatebox{-45}{\makebox(0,0)[l]{\strut{}01/23}}}%
      \put(2068,818){\rotatebox{-45}{\makebox(0,0)[l]{\strut{}02/06}}}%
      \put(2353,818){\rotatebox{-45}{\makebox(0,0)[l]{\strut{}02/20}}}%
      \put(2639,818){\rotatebox{-45}{\makebox(0,0)[l]{\strut{}03/05}}}%
      \put(2925,818){\rotatebox{-45}{\makebox(0,0)[l]{\strut{}03/19}}}%
      \put(3211,818){\rotatebox{-45}{\makebox(0,0)[l]{\strut{}04/02}}}%
      \put(3343,1013){\makebox(0,0)[l]{\strut{}$0$}}%
      \put(3343,1238){\makebox(0,0)[l]{\strut{}$2000$}}%
      \put(3343,1463){\makebox(0,0)[l]{\strut{}$4000$}}%
      \put(3343,1688){\makebox(0,0)[l]{\strut{}$6000$}}%
      \put(3343,1913){\makebox(0,0)[l]{\strut{}$8000$}}%
      \put(3343,2138){\makebox(0,0)[l]{\strut{}$10000$}}%
      \put(3343,2363){\makebox(0,0)[l]{\strut{}$12000$}}%
      \put(3343,2588){\makebox(0,0)[l]{\strut{}$14000$}}%
      \put(3343,2813){\makebox(0,0)[l]{\strut{}$16000$}}%
      \put(3343,3038){\makebox(0,0)[l]{\strut{}$18000$}}%
      \put(3343,3263){\makebox(0,0)[l]{\strut{}$20000$}}%
    }%
    \gplgaddtomacro\gplfronttext{%
      \csname LTb\endcsname
      \put(209,2138){\rotatebox{-270}{\makebox(0,0){\strut{}Confirmed Cases}}}%
      \put(4245,2138){\rotatebox{-270}{\makebox(0,0){\strut{}Article Count}}}%
      \put(2210,154){\makebox(0,0){\strut{}Day and Month (2020)}}%
      \csname LTb\endcsname
      \put(2530,3090){\makebox(0,0)[r]{\strut{}Confirmed}}%
      \csname LTb\endcsname
      \put(2530,2870){\makebox(0,0)[r]{\strut{}Article}}%
    }%
    \gplbacktext
    \put(0,0){\includegraphics{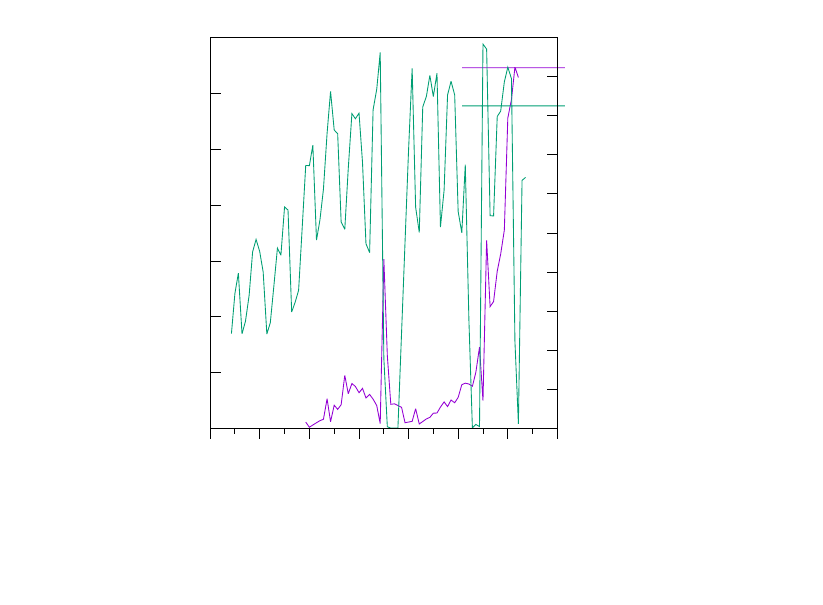}}%
    \gplfronttext
  \end{picture}%
\endgroup

%% file: plots/cn_filter.tex
\begingroup
  \makeatletter
  \providecommand\color[2][]{%
    \GenericError{(gnuplot) \space\space\space\@spaces}{%
      Package color not loaded in conjunction with
      terminal option `colourtext'%
    }{See the gnuplot documentation for explanation.%
    }{Either use 'blacktext' in gnuplot or load the package
      color.sty in LaTeX.}%
    \renewcommand\color[2][]{}%
  }%
  \providecommand\includegraphics[2][]{%
    \GenericError{(gnuplot) \space\space\space\@spaces}{%
      Package graphicx or graphics not loaded%
    }{See the gnuplot documentation for explanation.%
    }{The gnuplot epslatex terminal needs graphicx.sty or graphics.sty.}%
    \renewcommand\includegraphics[2][]{}%
  }%
  \providecommand\rotatebox[2]{#2}%
  \@ifundefined{ifGPcolor}{%
    \newif\ifGPcolor
    \GPcolortrue
  }{}%
  \@ifundefined{ifGPblacktext}{%
    \newif\ifGPblacktext
    \GPblacktextfalse
  }{}%
  \let\gplgaddtomacro\g@addto@macro
  \gdef\gplbacktext{}%
  \gdef\gplfronttext{}%
  \makeatother
  \ifGPblacktext
    \def\colorrgb#1{}%
    \def\colorgray#1{}%
  \else
    \ifGPcolor
      \def\colorrgb#1{\color[rgb]{#1}}%
      \def\colorgray#1{\color[gray]{#1}}%
      \expandafter\def\csname LTw\endcsname{\color{white}}%
      \expandafter\def\csname LTb\endcsname{\color{black}}%
      \expandafter\def\csname LTa\endcsname{\color{black}}%
      \expandafter\def\csname LT0\endcsname{\color[rgb]{1,0,0}}%
      \expandafter\def\csname LT1\endcsname{\color[rgb]{0,1,0}}%
      \expandafter\def\csname LT2\endcsname{\color[rgb]{0,0,1}}%
      \expandafter\def\csname LT3\endcsname{\color[rgb]{1,0,1}}%
      \expandafter\def\csname LT4\endcsname{\color[rgb]{0,1,1}}%
      \expandafter\def\csname LT5\endcsname{\color[rgb]{1,1,0}}%
      \expandafter\def\csname LT6\endcsname{\color[rgb]{0,0,0}}%
      \expandafter\def\csname LT7\endcsname{\color[rgb]{1,0.3,0}}%
      \expandafter\def\csname LT8\endcsname{\color[rgb]{0.5,0.5,0.5}}%
    \else
      \def\colorrgb#1{\color{black}}%
      \def\colorgray#1{\color[gray]{#1}}%
      \expandafter\def\csname LTw\endcsname{\color{white}}%
      \expandafter\def\csname LTb\endcsname{\color{black}}%
      \expandafter\def\csname LTa\endcsname{\color{black}}%
      \expandafter\def\csname LT0\endcsname{\color{black}}%
      \expandafter\def\csname LT1\endcsname{\color{black}}%
      \expandafter\def\csname LT2\endcsname{\color{black}}%
      \expandafter\def\csname LT3\endcsname{\color{black}}%
      \expandafter\def\csname LT4\endcsname{\color{black}}%
      \expandafter\def\csname LT5\endcsname{\color{black}}%
      \expandafter\def\csname LT6\endcsname{\color{black}}%
      \expandafter\def\csname LT7\endcsname{\color{black}}%
      \expandafter\def\csname LT8\endcsname{\color{black}}%
    \fi
  \fi
    \setlength{\unitlength}{0.0500bp}%
    \ifx\gptboxheight\undefined%
      \newlength{\gptboxheight}%
      \newlength{\gptboxwidth}%
      \newsavebox{\gptboxtext}%
    \fi%
    \setlength{\fboxrule}{0.5pt}%
    \setlength{\fboxsep}{1pt}%
\begin{picture}(4752.00,3484.00)%
    \gplgaddtomacro\gplbacktext{%
      \csname LTb\endcsname
      \put(1078,1013){\makebox(0,0)[r]{\strut{}$0$}}%
      \put(1078,1294){\makebox(0,0)[r]{\strut{}$2000$}}%
      \put(1078,1576){\makebox(0,0)[r]{\strut{}$4000$}}%
      \put(1078,1857){\makebox(0,0)[r]{\strut{}$6000$}}%
      \put(1078,2138){\makebox(0,0)[r]{\strut{}$8000$}}%
      \put(1078,2419){\makebox(0,0)[r]{\strut{}$10000$}}%
      \put(1078,2701){\makebox(0,0)[r]{\strut{}$12000$}}%
      \put(1078,2982){\makebox(0,0)[r]{\strut{}$14000$}}%
      \put(1078,3263){\makebox(0,0)[r]{\strut{}$16000$}}%
      \put(1210,818){\rotatebox{-45}{\makebox(0,0)[l]{\strut{}01/09}}}%
      \put(1588,818){\rotatebox{-45}{\makebox(0,0)[l]{\strut{}01/23}}}%
      \put(1965,818){\rotatebox{-45}{\makebox(0,0)[l]{\strut{}02/06}}}%
      \put(2343,818){\rotatebox{-45}{\makebox(0,0)[l]{\strut{}02/20}}}%
      \put(2720,818){\rotatebox{-45}{\makebox(0,0)[l]{\strut{}03/05}}}%
      \put(3098,818){\rotatebox{-45}{\makebox(0,0)[l]{\strut{}03/19}}}%
      \put(3475,818){\rotatebox{-45}{\makebox(0,0)[l]{\strut{}04/02}}}%
      \put(3607,1013){\makebox(0,0)[l]{\strut{}$0$}}%
      \put(3607,1263){\makebox(0,0)[l]{\strut{}$50$}}%
      \put(3607,1513){\makebox(0,0)[l]{\strut{}$100$}}%
      \put(3607,1763){\makebox(0,0)[l]{\strut{}$150$}}%
      \put(3607,2013){\makebox(0,0)[l]{\strut{}$200$}}%
      \put(3607,2263){\makebox(0,0)[l]{\strut{}$250$}}%
      \put(3607,2513){\makebox(0,0)[l]{\strut{}$300$}}%
      \put(3607,2763){\makebox(0,0)[l]{\strut{}$350$}}%
      \put(3607,3013){\makebox(0,0)[l]{\strut{}$400$}}%
      \put(3607,3263){\makebox(0,0)[l]{\strut{}$450$}}%
    }%
    \gplgaddtomacro\gplfronttext{%
      \csname LTb\endcsname
      \put(209,2138){\rotatebox{-270}{\makebox(0,0){\strut{}Confirmed Cases}}}%
      \put(4245,2138){\rotatebox{-270}{\makebox(0,0){\strut{}Article Count}}}%
      \put(2342,154){\makebox(0,0){\strut{}Day and Month (2020)}}%
      \csname LTb\endcsname
      \put(2530,3090){\makebox(0,0)[r]{\strut{}Confirmed}}%
      \csname LTb\endcsname
      \put(2530,2870){\makebox(0,0)[r]{\strut{}Article}}%
    }%
    \gplbacktext
    \put(0,0){\includegraphics{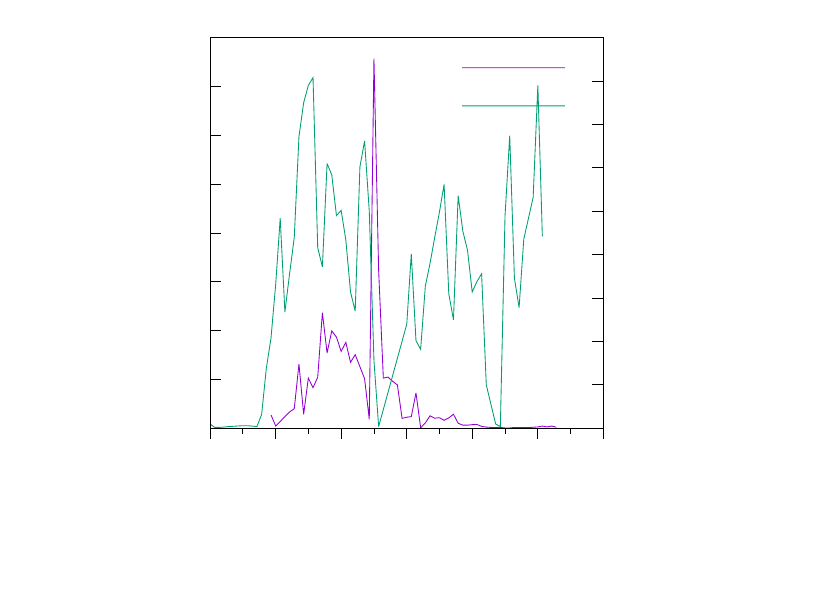}}%
    \gplfronttext
  \end{picture}%
\endgroup

%% file: plots/cn_nofilter.tex
\begingroup
  \makeatletter
  \providecommand\color[2][]{%
    \GenericError{(gnuplot) \space\space\space\@spaces}{%
      Package color not loaded in conjunction with
      terminal option `colourtext'%
    }{See the gnuplot documentation for explanation.%
    }{Either use 'blacktext' in gnuplot or load the package
      color.sty in LaTeX.}%
    \renewcommand\color[2][]{}%
  }%
  \providecommand\includegraphics[2][]{%
    \GenericError{(gnuplot) \space\space\space\@spaces}{%
      Package graphicx or graphics not loaded%
    }{See the gnuplot documentation for explanation.%
    }{The gnuplot epslatex terminal needs graphicx.sty or graphics.sty.}%
    \renewcommand\includegraphics[2][]{}%
  }%
  \providecommand\rotatebox[2]{#2}%
  \@ifundefined{ifGPcolor}{%
    \newif\ifGPcolor
    \GPcolortrue
  }{}%
  \@ifundefined{ifGPblacktext}{%
    \newif\ifGPblacktext
    \GPblacktextfalse
  }{}%
  \let\gplgaddtomacro\g@addto@macro
  \gdef\gplbacktext{}%
  \gdef\gplfronttext{}%
  \makeatother
  \ifGPblacktext
    \def\colorrgb#1{}%
    \def\colorgray#1{}%
  \else
    \ifGPcolor
      \def\colorrgb#1{\color[rgb]{#1}}%
      \def\colorgray#1{\color[gray]{#1}}%
      \expandafter\def\csname LTw\endcsname{\color{white}}%
      \expandafter\def\csname LTb\endcsname{\color{black}}%
      \expandafter\def\csname LTa\endcsname{\color{black}}%
      \expandafter\def\csname LT0\endcsname{\color[rgb]{1,0,0}}%
      \expandafter\def\csname LT1\endcsname{\color[rgb]{0,1,0}}%
      \expandafter\def\csname LT2\endcsname{\color[rgb]{0,0,1}}%
      \expandafter\def\csname LT3\endcsname{\color[rgb]{1,0,1}}%
      \expandafter\def\csname LT4\endcsname{\color[rgb]{0,1,1}}%
      \expandafter\def\csname LT5\endcsname{\color[rgb]{1,1,0}}%
      \expandafter\def\csname LT6\endcsname{\color[rgb]{0,0,0}}%
      \expandafter\def\csname LT7\endcsname{\color[rgb]{1,0.3,0}}%
      \expandafter\def\csname LT8\endcsname{\color[rgb]{0.5,0.5,0.5}}%
    \else
      \def\colorrgb#1{\color{black}}%
      \def\colorgray#1{\color[gray]{#1}}%
      \expandafter\def\csname LTw\endcsname{\color{white}}%
      \expandafter\def\csname LTb\endcsname{\color{black}}%
      \expandafter\def\csname LTa\endcsname{\color{black}}%
      \expandafter\def\csname LT0\endcsname{\color{black}}%
      \expandafter\def\csname LT1\endcsname{\color{black}}%
      \expandafter\def\csname LT2\endcsname{\color{black}}%
      \expandafter\def\csname LT3\endcsname{\color{black}}%
      \expandafter\def\csname LT4\endcsname{\color{black}}%
      \expandafter\def\csname LT5\endcsname{\color{black}}%
      \expandafter\def\csname LT6\endcsname{\color{black}}%
      \expandafter\def\csname LT7\endcsname{\color{black}}%
      \expandafter\def\csname LT8\endcsname{\color{black}}%
    \fi
  \fi
    \setlength{\unitlength}{0.0500bp}%
    \ifx\gptboxheight\undefined%
      \newlength{\gptboxheight}%
      \newlength{\gptboxwidth}%
      \newsavebox{\gptboxtext}%
    \fi%
    \setlength{\fboxrule}{0.5pt}%
    \setlength{\fboxsep}{1pt}%
\begin{picture}(4752.00,3484.00)%
    \gplgaddtomacro\gplbacktext{%
      \csname LTb\endcsname
      \put(1078,1013){\makebox(0,0)[r]{\strut{}$0$}}%
      \put(1078,1294){\makebox(0,0)[r]{\strut{}$2000$}}%
      \put(1078,1576){\makebox(0,0)[r]{\strut{}$4000$}}%
      \put(1078,1857){\makebox(0,0)[r]{\strut{}$6000$}}%
      \put(1078,2138){\makebox(0,0)[r]{\strut{}$8000$}}%
      \put(1078,2419){\makebox(0,0)[r]{\strut{}$10000$}}%
      \put(1078,2701){\makebox(0,0)[r]{\strut{}$12000$}}%
      \put(1078,2982){\makebox(0,0)[r]{\strut{}$14000$}}%
      \put(1078,3263){\makebox(0,0)[r]{\strut{}$16000$}}%
      \put(1210,818){\rotatebox{-45}{\makebox(0,0)[l]{\strut{}12/26}}}%
      \put(1534,818){\rotatebox{-45}{\makebox(0,0)[l]{\strut{}01/09}}}%
      \put(1857,818){\rotatebox{-45}{\makebox(0,0)[l]{\strut{}01/23}}}%
      \put(2181,818){\rotatebox{-45}{\makebox(0,0)[l]{\strut{}02/06}}}%
      \put(2504,818){\rotatebox{-45}{\makebox(0,0)[l]{\strut{}02/20}}}%
      \put(2828,818){\rotatebox{-45}{\makebox(0,0)[l]{\strut{}03/05}}}%
      \put(3151,818){\rotatebox{-45}{\makebox(0,0)[l]{\strut{}03/19}}}%
      \put(3475,818){\rotatebox{-45}{\makebox(0,0)[l]{\strut{}04/02}}}%
      \put(3607,1013){\makebox(0,0)[l]{\strut{}$0$}}%
      \put(3607,1263){\makebox(0,0)[l]{\strut{}$100$}}%
      \put(3607,1513){\makebox(0,0)[l]{\strut{}$200$}}%
      \put(3607,1763){\makebox(0,0)[l]{\strut{}$300$}}%
      \put(3607,2013){\makebox(0,0)[l]{\strut{}$400$}}%
      \put(3607,2263){\makebox(0,0)[l]{\strut{}$500$}}%
      \put(3607,2513){\makebox(0,0)[l]{\strut{}$600$}}%
      \put(3607,2763){\makebox(0,0)[l]{\strut{}$700$}}%
      \put(3607,3013){\makebox(0,0)[l]{\strut{}$800$}}%
      \put(3607,3263){\makebox(0,0)[l]{\strut{}$900$}}%
    }%
    \gplgaddtomacro\gplfronttext{%
      \csname LTb\endcsname
      \put(209,2138){\rotatebox{-270}{\makebox(0,0){\strut{}Confirmed Cases}}}%
      \put(4245,2138){\rotatebox{-270}{\makebox(0,0){\strut{}Article Count}}}%
      \put(2342,154){\makebox(0,0){\strut{}Day and Month (2020)}}%
      \csname LTb\endcsname
      \put(2530,3090){\makebox(0,0)[r]{\strut{}Confirmed}}%
      \csname LTb\endcsname
      \put(2530,2870){\makebox(0,0)[r]{\strut{}Article}}%
    }%
    \gplbacktext
    \put(0,0){\includegraphics{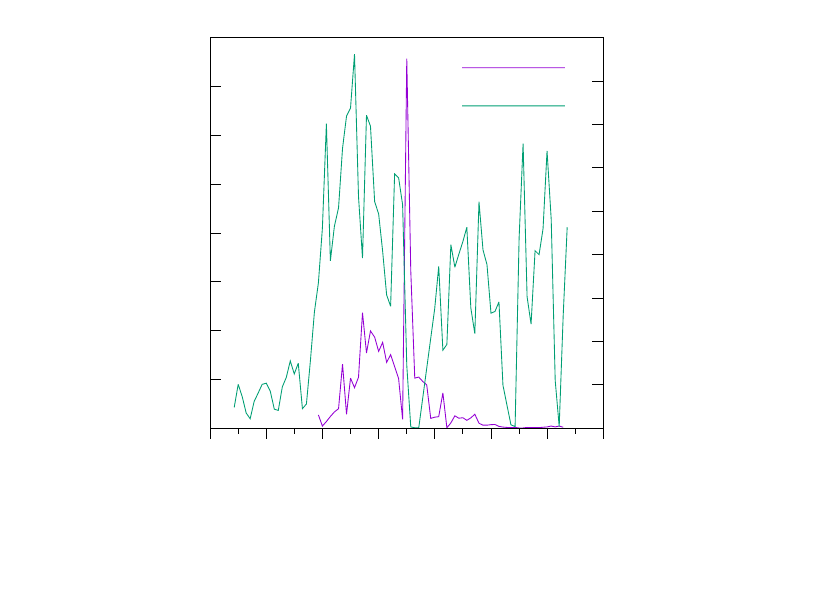}}%
    \gplfronttext
  \end{picture}%
\endgroup

%% file: plots/hubei_filter.tex
\begingroup
  \makeatletter
  \providecommand\color[2][]{%
    \GenericError{(gnuplot) \space\space\space\@spaces}{%
      Package color not loaded in conjunction with
      terminal option `colourtext'%
    }{See the gnuplot documentation for explanation.%
    }{Either use 'blacktext' in gnuplot or load the package
      color.sty in LaTeX.}%
    \renewcommand\color[2][]{}%
  }%
  \providecommand\includegraphics[2][]{%
    \GenericError{(gnuplot) \space\space\space\@spaces}{%
      Package graphicx or graphics not loaded%
    }{See the gnuplot documentation for explanation.%
    }{The gnuplot epslatex terminal needs graphicx.sty or graphics.sty.}%
    \renewcommand\includegraphics[2][]{}%
  }%
  \providecommand\rotatebox[2]{#2}%
  \@ifundefined{ifGPcolor}{%
    \newif\ifGPcolor
    \GPcolortrue
  }{}%
  \@ifundefined{ifGPblacktext}{%
    \newif\ifGPblacktext
    \GPblacktextfalse
  }{}%
  \let\gplgaddtomacro\g@addto@macro
  \gdef\gplbacktext{}%
  \gdef\gplfronttext{}%
  \makeatother
  \ifGPblacktext
    \def\colorrgb#1{}%
    \def\colorgray#1{}%
  \else
    \ifGPcolor
      \def\colorrgb#1{\color[rgb]{#1}}%
      \def\colorgray#1{\color[gray]{#1}}%
      \expandafter\def\csname LTw\endcsname{\color{white}}%
      \expandafter\def\csname LTb\endcsname{\color{black}}%
      \expandafter\def\csname LTa\endcsname{\color{black}}%
      \expandafter\def\csname LT0\endcsname{\color[rgb]{1,0,0}}%
      \expandafter\def\csname LT1\endcsname{\color[rgb]{0,1,0}}%
      \expandafter\def\csname LT2\endcsname{\color[rgb]{0,0,1}}%
      \expandafter\def\csname LT3\endcsname{\color[rgb]{1,0,1}}%
      \expandafter\def\csname LT4\endcsname{\color[rgb]{0,1,1}}%
      \expandafter\def\csname LT5\endcsname{\color[rgb]{1,1,0}}%
      \expandafter\def\csname LT6\endcsname{\color[rgb]{0,0,0}}%
      \expandafter\def\csname LT7\endcsname{\color[rgb]{1,0.3,0}}%
      \expandafter\def\csname LT8\endcsname{\color[rgb]{0.5,0.5,0.5}}%
    \else
      \def\colorrgb#1{\color{black}}%
      \def\colorgray#1{\color[gray]{#1}}%
      \expandafter\def\csname LTw\endcsname{\color{white}}%
      \expandafter\def\csname LTb\endcsname{\color{black}}%
      \expandafter\def\csname LTa\endcsname{\color{black}}%
      \expandafter\def\csname LT0\endcsname{\color{black}}%
      \expandafter\def\csname LT1\endcsname{\color{black}}%
      \expandafter\def\csname LT2\endcsname{\color{black}}%
      \expandafter\def\csname LT3\endcsname{\color{black}}%
      \expandafter\def\csname LT4\endcsname{\color{black}}%
      \expandafter\def\csname LT5\endcsname{\color{black}}%
      \expandafter\def\csname LT6\endcsname{\color{black}}%
      \expandafter\def\csname LT7\endcsname{\color{black}}%
      \expandafter\def\csname LT8\endcsname{\color{black}}%
    \fi
  \fi
    \setlength{\unitlength}{0.0500bp}%
    \ifx\gptboxheight\undefined%
      \newlength{\gptboxheight}%
      \newlength{\gptboxwidth}%
      \newsavebox{\gptboxtext}%
    \fi%
    \setlength{\fboxrule}{0.5pt}%
    \setlength{\fboxsep}{1pt}%
\begin{picture}(4752.00,3484.00)%
    \gplgaddtomacro\gplbacktext{%
      \csname LTb\endcsname
      \put(1078,1013){\makebox(0,0)[r]{\strut{}$0$}}%
      \put(1078,1294){\makebox(0,0)[r]{\strut{}$2000$}}%
      \put(1078,1576){\makebox(0,0)[r]{\strut{}$4000$}}%
      \put(1078,1857){\makebox(0,0)[r]{\strut{}$6000$}}%
      \put(1078,2138){\makebox(0,0)[r]{\strut{}$8000$}}%
      \put(1078,2419){\makebox(0,0)[r]{\strut{}$10000$}}%
      \put(1078,2701){\makebox(0,0)[r]{\strut{}$12000$}}%
      \put(1078,2982){\makebox(0,0)[r]{\strut{}$14000$}}%
      \put(1078,3263){\makebox(0,0)[r]{\strut{}$16000$}}%
      \put(1210,818){\rotatebox{-45}{\makebox(0,0)[l]{\strut{}01/09}}}%
      \put(1588,818){\rotatebox{-45}{\makebox(0,0)[l]{\strut{}01/23}}}%
      \put(1965,818){\rotatebox{-45}{\makebox(0,0)[l]{\strut{}02/06}}}%
      \put(2343,818){\rotatebox{-45}{\makebox(0,0)[l]{\strut{}02/20}}}%
      \put(2720,818){\rotatebox{-45}{\makebox(0,0)[l]{\strut{}03/05}}}%
      \put(3098,818){\rotatebox{-45}{\makebox(0,0)[l]{\strut{}03/19}}}%
      \put(3475,818){\rotatebox{-45}{\makebox(0,0)[l]{\strut{}04/02}}}%
      \put(3607,1013){\makebox(0,0)[l]{\strut{}$0$}}%
      \put(3607,1388){\makebox(0,0)[l]{\strut{}$50$}}%
      \put(3607,1763){\makebox(0,0)[l]{\strut{}$100$}}%
      \put(3607,2138){\makebox(0,0)[l]{\strut{}$150$}}%
      \put(3607,2513){\makebox(0,0)[l]{\strut{}$200$}}%
      \put(3607,2888){\makebox(0,0)[l]{\strut{}$250$}}%
      \put(3607,3263){\makebox(0,0)[l]{\strut{}$300$}}%
    }%
    \gplgaddtomacro\gplfronttext{%
      \csname LTb\endcsname
      \put(209,2138){\rotatebox{-270}{\makebox(0,0){\strut{}Confirmed Cases}}}%
      \put(4245,2138){\rotatebox{-270}{\makebox(0,0){\strut{}Article Count}}}%
      \put(2342,154){\makebox(0,0){\strut{}Day and Month (2020)}}%
      \csname LTb\endcsname
      \put(2530,3090){\makebox(0,0)[r]{\strut{}Confirmed}}%
      \csname LTb\endcsname
      \put(2530,2870){\makebox(0,0)[r]{\strut{}Article}}%
    }%
    \gplbacktext
    \put(0,0){\includegraphics{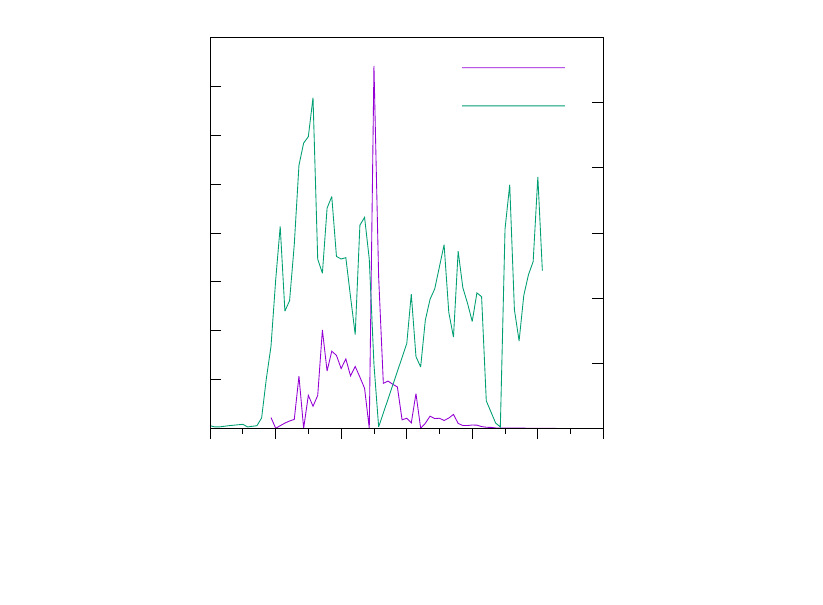}}%
    \gplfronttext
  \end{picture}%
\endgroup

%% file: plots/hubei_nofilter.tex
\begingroup
  \makeatletter
  \providecommand\color[2][]{%
    \GenericError{(gnuplot) \space\space\space\@spaces}{%
      Package color not loaded in conjunction with
      terminal option `colourtext'%
    }{See the gnuplot documentation for explanation.%
    }{Either use 'blacktext' in gnuplot or load the package
      color.sty in LaTeX.}%
    \renewcommand\color[2][]{}%
  }%
  \providecommand\includegraphics[2][]{%
    \GenericError{(gnuplot) \space\space\space\@spaces}{%
      Package graphicx or graphics not loaded%
    }{See the gnuplot documentation for explanation.%
    }{The gnuplot epslatex terminal needs graphicx.sty or graphics.sty.}%
    \renewcommand\includegraphics[2][]{}%
  }%
  \providecommand\rotatebox[2]{#2}%
  \@ifundefined{ifGPcolor}{%
    \newif\ifGPcolor
    \GPcolortrue
  }{}%
  \@ifundefined{ifGPblacktext}{%
    \newif\ifGPblacktext
    \GPblacktextfalse
  }{}%
  \let\gplgaddtomacro\g@addto@macro
  \gdef\gplbacktext{}%
  \gdef\gplfronttext{}%
  \makeatother
  \ifGPblacktext
    \def\colorrgb#1{}%
    \def\colorgray#1{}%
  \else
    \ifGPcolor
      \def\colorrgb#1{\color[rgb]{#1}}%
      \def\colorgray#1{\color[gray]{#1}}%
      \expandafter\def\csname LTw\endcsname{\color{white}}%
      \expandafter\def\csname LTb\endcsname{\color{black}}%
      \expandafter\def\csname LTa\endcsname{\color{black}}%
      \expandafter\def\csname LT0\endcsname{\color[rgb]{1,0,0}}%
      \expandafter\def\csname LT1\endcsname{\color[rgb]{0,1,0}}%
      \expandafter\def\csname LT2\endcsname{\color[rgb]{0,0,1}}%
      \expandafter\def\csname LT3\endcsname{\color[rgb]{1,0,1}}%
      \expandafter\def\csname LT4\endcsname{\color[rgb]{0,1,1}}%
      \expandafter\def\csname LT5\endcsname{\color[rgb]{1,1,0}}%
      \expandafter\def\csname LT6\endcsname{\color[rgb]{0,0,0}}%
      \expandafter\def\csname LT7\endcsname{\color[rgb]{1,0.3,0}}%
      \expandafter\def\csname LT8\endcsname{\color[rgb]{0.5,0.5,0.5}}%
    \else
      \def\colorrgb#1{\color{black}}%
      \def\colorgray#1{\color[gray]{#1}}%
      \expandafter\def\csname LTw\endcsname{\color{white}}%
      \expandafter\def\csname LTb\endcsname{\color{black}}%
      \expandafter\def\csname LTa\endcsname{\color{black}}%
      \expandafter\def\csname LT0\endcsname{\color{black}}%
      \expandafter\def\csname LT1\endcsname{\color{black}}%
      \expandafter\def\csname LT2\endcsname{\color{black}}%
      \expandafter\def\csname LT3\endcsname{\color{black}}%
      \expandafter\def\csname LT4\endcsname{\color{black}}%
      \expandafter\def\csname LT5\endcsname{\color{black}}%
      \expandafter\def\csname LT6\endcsname{\color{black}}%
      \expandafter\def\csname LT7\endcsname{\color{black}}%
      \expandafter\def\csname LT8\endcsname{\color{black}}%
    \fi
  \fi
    \setlength{\unitlength}{0.0500bp}%
    \ifx\gptboxheight\undefined%
      \newlength{\gptboxheight}%
      \newlength{\gptboxwidth}%
      \newsavebox{\gptboxtext}%
    \fi%
    \setlength{\fboxrule}{0.5pt}%
    \setlength{\fboxsep}{1pt}%
\begin{picture}(4752.00,3484.00)%
    \gplgaddtomacro\gplbacktext{%
      \csname LTb\endcsname
      \put(1078,1013){\makebox(0,0)[r]{\strut{}$0$}}%
      \put(1078,1294){\makebox(0,0)[r]{\strut{}$2000$}}%
      \put(1078,1576){\makebox(0,0)[r]{\strut{}$4000$}}%
      \put(1078,1857){\makebox(0,0)[r]{\strut{}$6000$}}%
      \put(1078,2138){\makebox(0,0)[r]{\strut{}$8000$}}%
      \put(1078,2419){\makebox(0,0)[r]{\strut{}$10000$}}%
      \put(1078,2701){\makebox(0,0)[r]{\strut{}$12000$}}%
      \put(1078,2982){\makebox(0,0)[r]{\strut{}$14000$}}%
      \put(1078,3263){\makebox(0,0)[r]{\strut{}$16000$}}%
      \put(1210,818){\rotatebox{-45}{\makebox(0,0)[l]{\strut{}12/26}}}%
      \put(1534,818){\rotatebox{-45}{\makebox(0,0)[l]{\strut{}01/09}}}%
      \put(1857,818){\rotatebox{-45}{\makebox(0,0)[l]{\strut{}01/23}}}%
      \put(2181,818){\rotatebox{-45}{\makebox(0,0)[l]{\strut{}02/06}}}%
      \put(2504,818){\rotatebox{-45}{\makebox(0,0)[l]{\strut{}02/20}}}%
      \put(2828,818){\rotatebox{-45}{\makebox(0,0)[l]{\strut{}03/05}}}%
      \put(3151,818){\rotatebox{-45}{\makebox(0,0)[l]{\strut{}03/19}}}%
      \put(3475,818){\rotatebox{-45}{\makebox(0,0)[l]{\strut{}04/02}}}%
      \put(3607,1013){\makebox(0,0)[l]{\strut{}$0$}}%
      \put(3607,1388){\makebox(0,0)[l]{\strut{}$100$}}%
      \put(3607,1763){\makebox(0,0)[l]{\strut{}$200$}}%
      \put(3607,2138){\makebox(0,0)[l]{\strut{}$300$}}%
      \put(3607,2513){\makebox(0,0)[l]{\strut{}$400$}}%
      \put(3607,2888){\makebox(0,0)[l]{\strut{}$500$}}%
      \put(3607,3263){\makebox(0,0)[l]{\strut{}$600$}}%
    }%
    \gplgaddtomacro\gplfronttext{%
      \csname LTb\endcsname
      \put(209,2138){\rotatebox{-270}{\makebox(0,0){\strut{}Confirmed Cases}}}%
      \put(4245,2138){\rotatebox{-270}{\makebox(0,0){\strut{}Article Count}}}%
      \put(2342,154){\makebox(0,0){\strut{}Day and Month (2020)}}%
      \csname LTb\endcsname
      \put(2530,3090){\makebox(0,0)[r]{\strut{}Confirmed}}%
      \csname LTb\endcsname
      \put(2530,2870){\makebox(0,0)[r]{\strut{}Article}}%
    }%
    \gplbacktext
    \put(0,0){\includegraphics{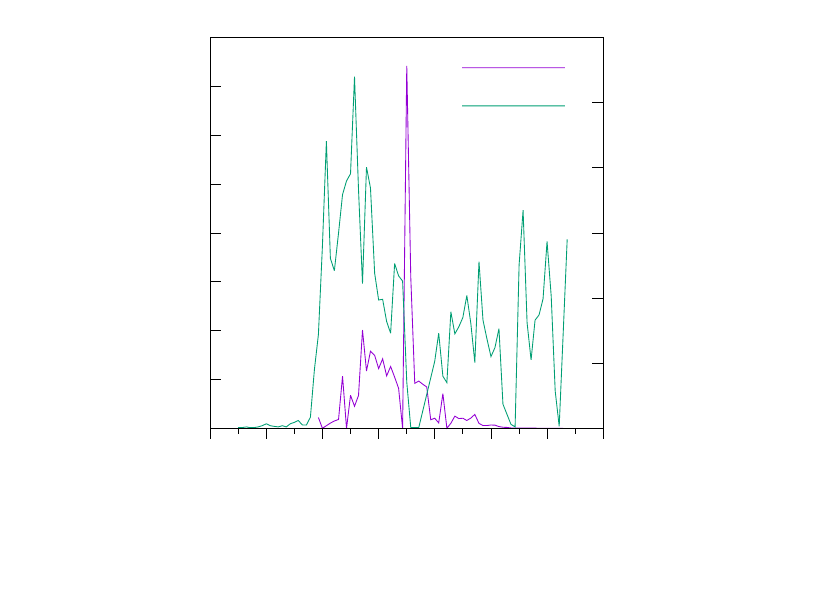}}%
    \gplfronttext
  \end{picture}%
\endgroup

%% file: plots/us_filter.tex
\begingroup
  \makeatletter
  \providecommand\color[2][]{%
    \GenericError{(gnuplot) \space\space\space\@spaces}{%
      Package color not loaded in conjunction with
      terminal option `colourtext'%
    }{See the gnuplot documentation for explanation.%
    }{Either use 'blacktext' in gnuplot or load the package
      color.sty in LaTeX.}%
    \renewcommand\color[2][]{}%
  }%
  \providecommand\includegraphics[2][]{%
    \GenericError{(gnuplot) \space\space\space\@spaces}{%
      Package graphicx or graphics not loaded%
    }{See the gnuplot documentation for explanation.%
    }{The gnuplot epslatex terminal needs graphicx.sty or graphics.sty.}%
    \renewcommand\includegraphics[2][]{}%
  }%
  \providecommand\rotatebox[2]{#2}%
  \@ifundefined{ifGPcolor}{%
    \newif\ifGPcolor
    \GPcolortrue
  }{}%
  \@ifundefined{ifGPblacktext}{%
    \newif\ifGPblacktext
    \GPblacktextfalse
  }{}%
  \let\gplgaddtomacro\g@addto@macro
  \gdef\gplbacktext{}%
  \gdef\gplfronttext{}%
  \makeatother
  \ifGPblacktext
    \def\colorrgb#1{}%
    \def\colorgray#1{}%
  \else
    \ifGPcolor
      \def\colorrgb#1{\color[rgb]{#1}}%
      \def\colorgray#1{\color[gray]{#1}}%
      \expandafter\def\csname LTw\endcsname{\color{white}}%
      \expandafter\def\csname LTb\endcsname{\color{black}}%
      \expandafter\def\csname LTa\endcsname{\color{black}}%
      \expandafter\def\csname LT0\endcsname{\color[rgb]{1,0,0}}%
      \expandafter\def\csname LT1\endcsname{\color[rgb]{0,1,0}}%
      \expandafter\def\csname LT2\endcsname{\color[rgb]{0,0,1}}%
      \expandafter\def\csname LT3\endcsname{\color[rgb]{1,0,1}}%
      \expandafter\def\csname LT4\endcsname{\color[rgb]{0,1,1}}%
      \expandafter\def\csname LT5\endcsname{\color[rgb]{1,1,0}}%
      \expandafter\def\csname LT6\endcsname{\color[rgb]{0,0,0}}%
      \expandafter\def\csname LT7\endcsname{\color[rgb]{1,0.3,0}}%
      \expandafter\def\csname LT8\endcsname{\color[rgb]{0.5,0.5,0.5}}%
    \else
      \def\colorrgb#1{\color{black}}%
      \def\colorgray#1{\color[gray]{#1}}%
      \expandafter\def\csname LTw\endcsname{\color{white}}%
      \expandafter\def\csname LTb\endcsname{\color{black}}%
      \expandafter\def\csname LTa\endcsname{\color{black}}%
      \expandafter\def\csname LT0\endcsname{\color{black}}%
      \expandafter\def\csname LT1\endcsname{\color{black}}%
      \expandafter\def\csname LT2\endcsname{\color{black}}%
      \expandafter\def\csname LT3\endcsname{\color{black}}%
      \expandafter\def\csname LT4\endcsname{\color{black}}%
      \expandafter\def\csname LT5\endcsname{\color{black}}%
      \expandafter\def\csname LT6\endcsname{\color{black}}%
      \expandafter\def\csname LT7\endcsname{\color{black}}%
      \expandafter\def\csname LT8\endcsname{\color{black}}%
    \fi
  \fi
    \setlength{\unitlength}{0.0500bp}%
    \ifx\gptboxheight\undefined%
      \newlength{\gptboxheight}%
      \newlength{\gptboxwidth}%
      \newsavebox{\gptboxtext}%
    \fi%
    \setlength{\fboxrule}{0.5pt}%
    \setlength{\fboxsep}{1pt}%
\begin{picture}(4752.00,3484.00)%
    \gplgaddtomacro\gplbacktext{%
      \csname LTb\endcsname
      \put(946,1013){\makebox(0,0)[r]{\strut{}$0$}}%
      \put(946,1294){\makebox(0,0)[r]{\strut{}$1000$}}%
      \put(946,1576){\makebox(0,0)[r]{\strut{}$2000$}}%
      \put(946,1857){\makebox(0,0)[r]{\strut{}$3000$}}%
      \put(946,2138){\makebox(0,0)[r]{\strut{}$4000$}}%
      \put(946,2419){\makebox(0,0)[r]{\strut{}$5000$}}%
      \put(946,2701){\makebox(0,0)[r]{\strut{}$6000$}}%
      \put(946,2982){\makebox(0,0)[r]{\strut{}$7000$}}%
      \put(946,3263){\makebox(0,0)[r]{\strut{}$8000$}}%
      \put(1078,818){\rotatebox{-45}{\makebox(0,0)[l]{\strut{}01/09}}}%
      \put(1456,818){\rotatebox{-45}{\makebox(0,0)[l]{\strut{}01/23}}}%
      \put(1833,818){\rotatebox{-45}{\makebox(0,0)[l]{\strut{}02/06}}}%
      \put(2211,818){\rotatebox{-45}{\makebox(0,0)[l]{\strut{}02/20}}}%
      \put(2588,818){\rotatebox{-45}{\makebox(0,0)[l]{\strut{}03/05}}}%
      \put(2966,818){\rotatebox{-45}{\makebox(0,0)[l]{\strut{}03/19}}}%
      \put(3343,818){\rotatebox{-45}{\makebox(0,0)[l]{\strut{}04/02}}}%
      \put(3475,1013){\makebox(0,0)[l]{\strut{}$0$}}%
      \put(3475,1294){\makebox(0,0)[l]{\strut{}$500$}}%
      \put(3475,1576){\makebox(0,0)[l]{\strut{}$1000$}}%
      \put(3475,1857){\makebox(0,0)[l]{\strut{}$1500$}}%
      \put(3475,2138){\makebox(0,0)[l]{\strut{}$2000$}}%
      \put(3475,2419){\makebox(0,0)[l]{\strut{}$2500$}}%
      \put(3475,2701){\makebox(0,0)[l]{\strut{}$3000$}}%
      \put(3475,2982){\makebox(0,0)[l]{\strut{}$3500$}}%
      \put(3475,3263){\makebox(0,0)[l]{\strut{}$4000$}}%
    }%
    \gplgaddtomacro\gplfronttext{%
      \csname LTb\endcsname
      \put(209,2138){\rotatebox{-270}{\makebox(0,0){\strut{}Confirmed Cases}}}%
      \put(4245,2138){\rotatebox{-270}{\makebox(0,0){\strut{}Article Count}}}%
      \put(2210,154){\makebox(0,0){\strut{}Day and Month (2020)}}%
      \csname LTb\endcsname
      \put(2398,3090){\makebox(0,0)[r]{\strut{}Confirmed}}%
      \csname LTb\endcsname
      \put(2398,2870){\makebox(0,0)[r]{\strut{}Article}}%
    }%
    \gplbacktext
    \put(0,0){\includegraphics{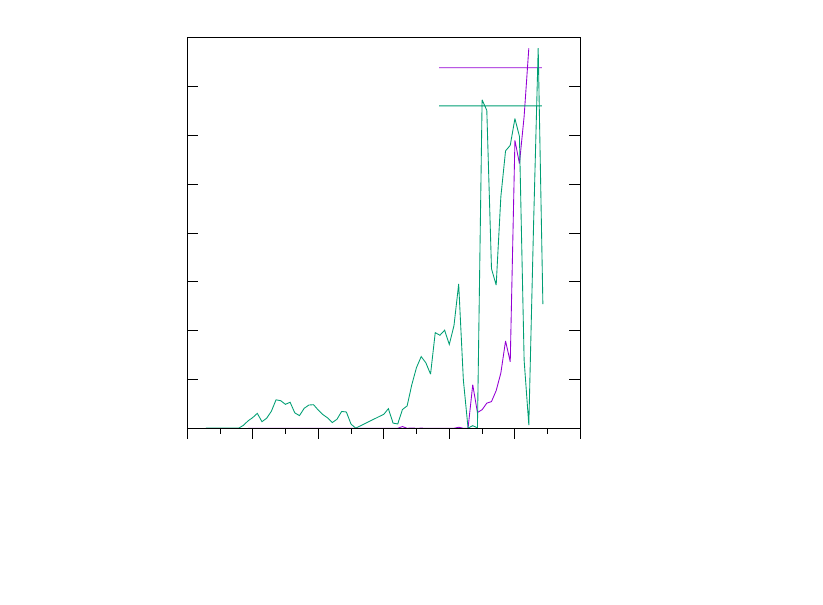}}%
    \gplfronttext
  \end{picture}%
\endgroup

%% file: plots/us_nofilter.tex
\begingroup
  \makeatletter
  \providecommand\color[2][]{%
    \GenericError{(gnuplot) \space\space\space\@spaces}{%
      Package color not loaded in conjunction with
      terminal option `colourtext'%
    }{See the gnuplot documentation for explanation.%
    }{Either use 'blacktext' in gnuplot or load the package
      color.sty in LaTeX.}%
    \renewcommand\color[2][]{}%
  }%
  \providecommand\includegraphics[2][]{%
    \GenericError{(gnuplot) \space\space\space\@spaces}{%
      Package graphicx or graphics not loaded%
    }{See the gnuplot documentation for explanation.%
    }{The gnuplot epslatex terminal needs graphicx.sty or graphics.sty.}%
    \renewcommand\includegraphics[2][]{}%
  }%
  \providecommand\rotatebox[2]{#2}%
  \@ifundefined{ifGPcolor}{%
    \newif\ifGPcolor
    \GPcolortrue
  }{}%
  \@ifundefined{ifGPblacktext}{%
    \newif\ifGPblacktext
    \GPblacktextfalse
  }{}%
  \let\gplgaddtomacro\g@addto@macro
  \gdef\gplbacktext{}%
  \gdef\gplfronttext{}%
  \makeatother
  \ifGPblacktext
    \def\colorrgb#1{}%
    \def\colorgray#1{}%
  \else
    \ifGPcolor
      \def\colorrgb#1{\color[rgb]{#1}}%
      \def\colorgray#1{\color[gray]{#1}}%
      \expandafter\def\csname LTw\endcsname{\color{white}}%
      \expandafter\def\csname LTb\endcsname{\color{black}}%
      \expandafter\def\csname LTa\endcsname{\color{black}}%
      \expandafter\def\csname LT0\endcsname{\color[rgb]{1,0,0}}%
      \expandafter\def\csname LT1\endcsname{\color[rgb]{0,1,0}}%
      \expandafter\def\csname LT2\endcsname{\color[rgb]{0,0,1}}%
      \expandafter\def\csname LT3\endcsname{\color[rgb]{1,0,1}}%
      \expandafter\def\csname LT4\endcsname{\color[rgb]{0,1,1}}%
      \expandafter\def\csname LT5\endcsname{\color[rgb]{1,1,0}}%
      \expandafter\def\csname LT6\endcsname{\color[rgb]{0,0,0}}%
      \expandafter\def\csname LT7\endcsname{\color[rgb]{1,0.3,0}}%
      \expandafter\def\csname LT8\endcsname{\color[rgb]{0.5,0.5,0.5}}%
    \else
      \def\colorrgb#1{\color{black}}%
      \def\colorgray#1{\color[gray]{#1}}%
      \expandafter\def\csname LTw\endcsname{\color{white}}%
      \expandafter\def\csname LTb\endcsname{\color{black}}%
      \expandafter\def\csname LTa\endcsname{\color{black}}%
      \expandafter\def\csname LT0\endcsname{\color{black}}%
      \expandafter\def\csname LT1\endcsname{\color{black}}%
      \expandafter\def\csname LT2\endcsname{\color{black}}%
      \expandafter\def\csname LT3\endcsname{\color{black}}%
      \expandafter\def\csname LT4\endcsname{\color{black}}%
      \expandafter\def\csname LT5\endcsname{\color{black}}%
      \expandafter\def\csname LT6\endcsname{\color{black}}%
      \expandafter\def\csname LT7\endcsname{\color{black}}%
      \expandafter\def\csname LT8\endcsname{\color{black}}%
    \fi
  \fi
    \setlength{\unitlength}{0.0500bp}%
    \ifx\gptboxheight\undefined%
      \newlength{\gptboxheight}%
      \newlength{\gptboxwidth}%
      \newsavebox{\gptboxtext}%
    \fi%
    \setlength{\fboxrule}{0.5pt}%
    \setlength{\fboxsep}{1pt}%
\begin{picture}(4752.00,3484.00)%
    \gplgaddtomacro\gplbacktext{%
      \csname LTb\endcsname
      \put(946,1013){\makebox(0,0)[r]{\strut{}$0$}}%
      \put(946,1294){\makebox(0,0)[r]{\strut{}$1000$}}%
      \put(946,1576){\makebox(0,0)[r]{\strut{}$2000$}}%
      \put(946,1857){\makebox(0,0)[r]{\strut{}$3000$}}%
      \put(946,2138){\makebox(0,0)[r]{\strut{}$4000$}}%
      \put(946,2419){\makebox(0,0)[r]{\strut{}$5000$}}%
      \put(946,2701){\makebox(0,0)[r]{\strut{}$6000$}}%
      \put(946,2982){\makebox(0,0)[r]{\strut{}$7000$}}%
      \put(946,3263){\makebox(0,0)[r]{\strut{}$8000$}}%
      \put(1078,818){\rotatebox{-45}{\makebox(0,0)[l]{\strut{}12/26}}}%
      \put(1383,818){\rotatebox{-45}{\makebox(0,0)[l]{\strut{}01/09}}}%
      \put(1687,818){\rotatebox{-45}{\makebox(0,0)[l]{\strut{}01/23}}}%
      \put(1992,818){\rotatebox{-45}{\makebox(0,0)[l]{\strut{}02/06}}}%
      \put(2297,818){\rotatebox{-45}{\makebox(0,0)[l]{\strut{}02/20}}}%
      \put(2602,818){\rotatebox{-45}{\makebox(0,0)[l]{\strut{}03/05}}}%
      \put(2906,818){\rotatebox{-45}{\makebox(0,0)[l]{\strut{}03/19}}}%
      \put(3211,818){\rotatebox{-45}{\makebox(0,0)[l]{\strut{}04/02}}}%
      \put(3343,1013){\makebox(0,0)[l]{\strut{}$0$}}%
      \put(3343,1388){\makebox(0,0)[l]{\strut{}$2000$}}%
      \put(3343,1763){\makebox(0,0)[l]{\strut{}$4000$}}%
      \put(3343,2138){\makebox(0,0)[l]{\strut{}$6000$}}%
      \put(3343,2513){\makebox(0,0)[l]{\strut{}$8000$}}%
      \put(3343,2888){\makebox(0,0)[l]{\strut{}$10000$}}%
      \put(3343,3263){\makebox(0,0)[l]{\strut{}$12000$}}%
    }%
    \gplgaddtomacro\gplfronttext{%
      \csname LTb\endcsname
      \put(209,2138){\rotatebox{-270}{\makebox(0,0){\strut{}Confirmed Cases}}}%
      \put(4245,2138){\rotatebox{-270}{\makebox(0,0){\strut{}Article Count}}}%
      \put(2144,154){\makebox(0,0){\strut{}Day and Month (2020)}}%
      \csname LTb\endcsname
      \put(2398,3090){\makebox(0,0)[r]{\strut{}Confirmed}}%
      \csname LTb\endcsname
      \put(2398,2870){\makebox(0,0)[r]{\strut{}Article}}%
    }%
    \gplbacktext
    \put(0,0){\includegraphics{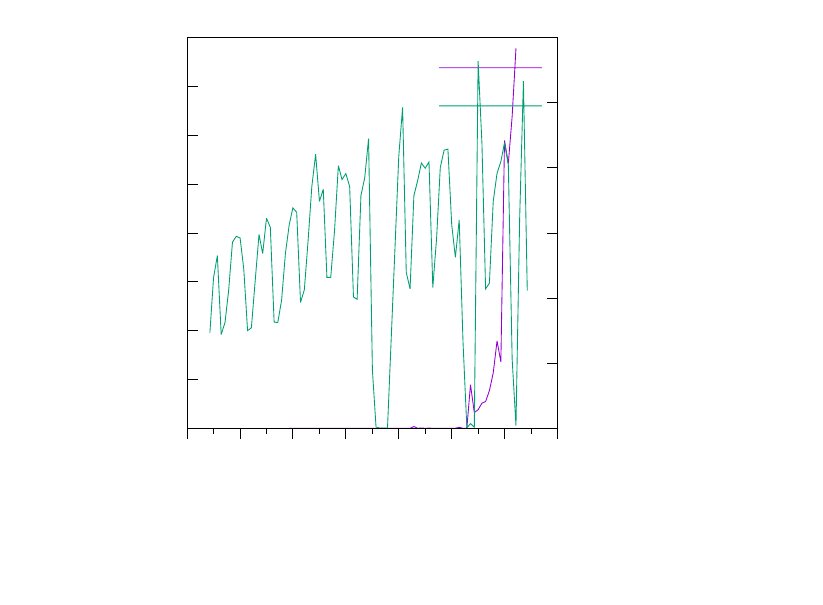}}%
    \gplfronttext
  \end{picture}%
\endgroup

%% file: plots/washington_filter.tex
\begingroup
  \makeatletter
  \providecommand\color[2][]{%
    \GenericError{(gnuplot) \space\space\space\@spaces}{%
      Package color not loaded in conjunction with
      terminal option `colourtext'%
    }{See the gnuplot documentation for explanation.%
    }{Either use 'blacktext' in gnuplot or load the package
      color.sty in LaTeX.}%
    \renewcommand\color[2][]{}%
  }%
  \providecommand\includegraphics[2][]{%
    \GenericError{(gnuplot) \space\space\space\@spaces}{%
      Package graphicx or graphics not loaded%
    }{See the gnuplot documentation for explanation.%
    }{The gnuplot epslatex terminal needs graphicx.sty or graphics.sty.}%
    \renewcommand\includegraphics[2][]{}%
  }%
  \providecommand\rotatebox[2]{#2}%
  \@ifundefined{ifGPcolor}{%
    \newif\ifGPcolor
    \GPcolortrue
  }{}%
  \@ifundefined{ifGPblacktext}{%
    \newif\ifGPblacktext
    \GPblacktextfalse
  }{}%
  \let\gplgaddtomacro\g@addto@macro
  \gdef\gplbacktext{}%
  \gdef\gplfronttext{}%
  \makeatother
  \ifGPblacktext
    \def\colorrgb#1{}%
    \def\colorgray#1{}%
  \else
    \ifGPcolor
      \def\colorrgb#1{\color[rgb]{#1}}%
      \def\colorgray#1{\color[gray]{#1}}%
      \expandafter\def\csname LTw\endcsname{\color{white}}%
      \expandafter\def\csname LTb\endcsname{\color{black}}%
      \expandafter\def\csname LTa\endcsname{\color{black}}%
      \expandafter\def\csname LT0\endcsname{\color[rgb]{1,0,0}}%
      \expandafter\def\csname LT1\endcsname{\color[rgb]{0,1,0}}%
      \expandafter\def\csname LT2\endcsname{\color[rgb]{0,0,1}}%
      \expandafter\def\csname LT3\endcsname{\color[rgb]{1,0,1}}%
      \expandafter\def\csname LT4\endcsname{\color[rgb]{0,1,1}}%
      \expandafter\def\csname LT5\endcsname{\color[rgb]{1,1,0}}%
      \expandafter\def\csname LT6\endcsname{\color[rgb]{0,0,0}}%
      \expandafter\def\csname LT7\endcsname{\color[rgb]{1,0.3,0}}%
      \expandafter\def\csname LT8\endcsname{\color[rgb]{0.5,0.5,0.5}}%
    \else
      \def\colorrgb#1{\color{black}}%
      \def\colorgray#1{\color[gray]{#1}}%
      \expandafter\def\csname LTw\endcsname{\color{white}}%
      \expandafter\def\csname LTb\endcsname{\color{black}}%
      \expandafter\def\csname LTa\endcsname{\color{black}}%
      \expandafter\def\csname LT0\endcsname{\color{black}}%
      \expandafter\def\csname LT1\endcsname{\color{black}}%
      \expandafter\def\csname LT2\endcsname{\color{black}}%
      \expandafter\def\csname LT3\endcsname{\color{black}}%
      \expandafter\def\csname LT4\endcsname{\color{black}}%
      \expandafter\def\csname LT5\endcsname{\color{black}}%
      \expandafter\def\csname LT6\endcsname{\color{black}}%
      \expandafter\def\csname LT7\endcsname{\color{black}}%
      \expandafter\def\csname LT8\endcsname{\color{black}}%
    \fi
  \fi
    \setlength{\unitlength}{0.0500bp}%
    \ifx\gptboxheight\undefined%
      \newlength{\gptboxheight}%
      \newlength{\gptboxwidth}%
      \newsavebox{\gptboxtext}%
    \fi%
    \setlength{\fboxrule}{0.5pt}%
    \setlength{\fboxsep}{1pt}%
\begin{picture}(4752.00,3484.00)%
    \gplgaddtomacro\gplbacktext{%
      \csname LTb\endcsname
      \put(814,1013){\makebox(0,0)[r]{\strut{}$0$}}%
      \put(814,1294){\makebox(0,0)[r]{\strut{}$50$}}%
      \put(814,1576){\makebox(0,0)[r]{\strut{}$100$}}%
      \put(814,1857){\makebox(0,0)[r]{\strut{}$150$}}%
      \put(814,2138){\makebox(0,0)[r]{\strut{}$200$}}%
      \put(814,2419){\makebox(0,0)[r]{\strut{}$250$}}%
      \put(814,2701){\makebox(0,0)[r]{\strut{}$300$}}%
      \put(814,2982){\makebox(0,0)[r]{\strut{}$350$}}%
      \put(814,3263){\makebox(0,0)[r]{\strut{}$400$}}%
      \put(946,818){\rotatebox{-45}{\makebox(0,0)[l]{\strut{}01/16}}}%
      \put(1199,818){\rotatebox{-45}{\makebox(0,0)[l]{\strut{}01/23}}}%
      \put(1452,818){\rotatebox{-45}{\makebox(0,0)[l]{\strut{}01/30}}}%
      \put(1705,818){\rotatebox{-45}{\makebox(0,0)[l]{\strut{}02/06}}}%
      \put(1958,818){\rotatebox{-45}{\makebox(0,0)[l]{\strut{}02/13}}}%
      \put(2211,818){\rotatebox{-45}{\makebox(0,0)[l]{\strut{}02/20}}}%
      \put(2463,818){\rotatebox{-45}{\makebox(0,0)[l]{\strut{}02/27}}}%
      \put(2716,818){\rotatebox{-45}{\makebox(0,0)[l]{\strut{}03/05}}}%
      \put(2969,818){\rotatebox{-45}{\makebox(0,0)[l]{\strut{}03/12}}}%
      \put(3222,818){\rotatebox{-45}{\makebox(0,0)[l]{\strut{}03/19}}}%
      \put(3475,818){\rotatebox{-45}{\makebox(0,0)[l]{\strut{}03/26}}}%
      \put(3607,1013){\makebox(0,0)[l]{\strut{}$0$}}%
      \put(3607,1263){\makebox(0,0)[l]{\strut{}$20$}}%
      \put(3607,1513){\makebox(0,0)[l]{\strut{}$40$}}%
      \put(3607,1763){\makebox(0,0)[l]{\strut{}$60$}}%
      \put(3607,2013){\makebox(0,0)[l]{\strut{}$80$}}%
      \put(3607,2263){\makebox(0,0)[l]{\strut{}$100$}}%
      \put(3607,2513){\makebox(0,0)[l]{\strut{}$120$}}%
      \put(3607,2763){\makebox(0,0)[l]{\strut{}$140$}}%
      \put(3607,3013){\makebox(0,0)[l]{\strut{}$160$}}%
      \put(3607,3263){\makebox(0,0)[l]{\strut{}$180$}}%
    }%
    \gplgaddtomacro\gplfronttext{%
      \csname LTb\endcsname
      \put(209,2138){\rotatebox{-270}{\makebox(0,0){\strut{}Confirmed Cases}}}%
      \put(4245,2138){\rotatebox{-270}{\makebox(0,0){\strut{}Article Count}}}%
      \put(2210,154){\makebox(0,0){\strut{}Day and Month (2020)}}%
      \csname LTb\endcsname
      \put(2266,3090){\makebox(0,0)[r]{\strut{}Confirmed}}%
      \csname LTb\endcsname
      \put(2266,2870){\makebox(0,0)[r]{\strut{}Article}}%
    }%
    \gplbacktext
    \put(0,0){\includegraphics{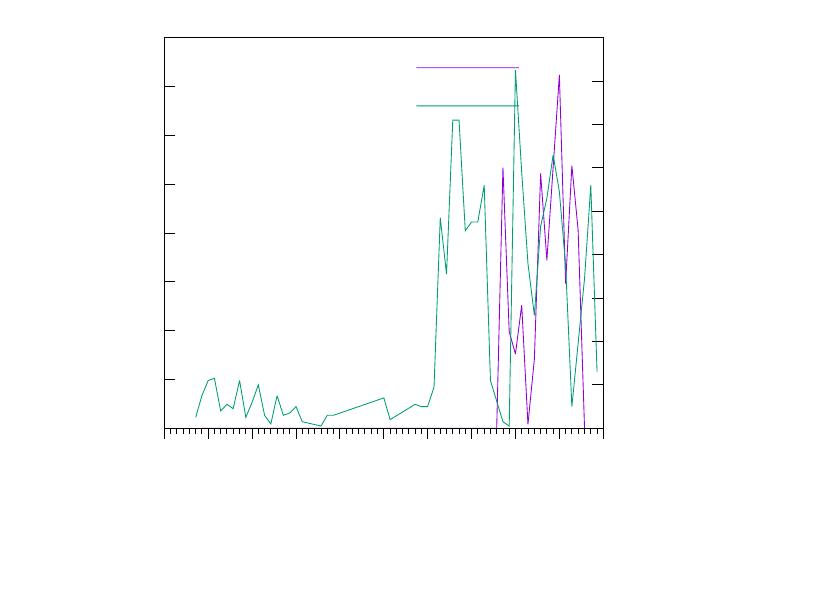}}%
    \gplfronttext
  \end{picture}%
\endgroup

%% file: plots/washington_nofilter.tex
\begingroup
  \makeatletter
  \providecommand\color[2][]{%
    \GenericError{(gnuplot) \space\space\space\@spaces}{%
      Package color not loaded in conjunction with
      terminal option `colourtext'%
    }{See the gnuplot documentation for explanation.%
    }{Either use 'blacktext' in gnuplot or load the package
      color.sty in LaTeX.}%
    \renewcommand\color[2][]{}%
  }%
  \providecommand\includegraphics[2][]{%
    \GenericError{(gnuplot) \space\space\space\@spaces}{%
      Package graphicx or graphics not loaded%
    }{See the gnuplot documentation for explanation.%
    }{The gnuplot epslatex terminal needs graphicx.sty or graphics.sty.}%
    \renewcommand\includegraphics[2][]{}%
  }%
  \providecommand\rotatebox[2]{#2}%
  \@ifundefined{ifGPcolor}{%
    \newif\ifGPcolor
    \GPcolortrue
  }{}%
  \@ifundefined{ifGPblacktext}{%
    \newif\ifGPblacktext
    \GPblacktextfalse
  }{}%
  \let\gplgaddtomacro\g@addto@macro
  \gdef\gplbacktext{}%
  \gdef\gplfronttext{}%
  \makeatother
  \ifGPblacktext
    \def\colorrgb#1{}%
    \def\colorgray#1{}%
  \else
    \ifGPcolor
      \def\colorrgb#1{\color[rgb]{#1}}%
      \def\colorgray#1{\color[gray]{#1}}%
      \expandafter\def\csname LTw\endcsname{\color{white}}%
      \expandafter\def\csname LTb\endcsname{\color{black}}%
      \expandafter\def\csname LTa\endcsname{\color{black}}%
      \expandafter\def\csname LT0\endcsname{\color[rgb]{1,0,0}}%
      \expandafter\def\csname LT1\endcsname{\color[rgb]{0,1,0}}%
      \expandafter\def\csname LT2\endcsname{\color[rgb]{0,0,1}}%
      \expandafter\def\csname LT3\endcsname{\color[rgb]{1,0,1}}%
      \expandafter\def\csname LT4\endcsname{\color[rgb]{0,1,1}}%
      \expandafter\def\csname LT5\endcsname{\color[rgb]{1,1,0}}%
      \expandafter\def\csname LT6\endcsname{\color[rgb]{0,0,0}}%
      \expandafter\def\csname LT7\endcsname{\color[rgb]{1,0.3,0}}%
      \expandafter\def\csname LT8\endcsname{\color[rgb]{0.5,0.5,0.5}}%
    \else
      \def\colorrgb#1{\color{black}}%
      \def\colorgray#1{\color[gray]{#1}}%
      \expandafter\def\csname LTw\endcsname{\color{white}}%
      \expandafter\def\csname LTb\endcsname{\color{black}}%
      \expandafter\def\csname LTa\endcsname{\color{black}}%
      \expandafter\def\csname LT0\endcsname{\color{black}}%
      \expandafter\def\csname LT1\endcsname{\color{black}}%
      \expandafter\def\csname LT2\endcsname{\color{black}}%
      \expandafter\def\csname LT3\endcsname{\color{black}}%
      \expandafter\def\csname LT4\endcsname{\color{black}}%
      \expandafter\def\csname LT5\endcsname{\color{black}}%
      \expandafter\def\csname LT6\endcsname{\color{black}}%
      \expandafter\def\csname LT7\endcsname{\color{black}}%
      \expandafter\def\csname LT8\endcsname{\color{black}}%
    \fi
  \fi
    \setlength{\unitlength}{0.0500bp}%
    \ifx\gptboxheight\undefined%
      \newlength{\gptboxheight}%
      \newlength{\gptboxwidth}%
      \newsavebox{\gptboxtext}%
    \fi%
    \setlength{\fboxrule}{0.5pt}%
    \setlength{\fboxsep}{1pt}%
\begin{picture}(4752.00,3484.00)%
    \gplgaddtomacro\gplbacktext{%
      \csname LTb\endcsname
      \put(814,1013){\makebox(0,0)[r]{\strut{}$0$}}%
      \put(814,1294){\makebox(0,0)[r]{\strut{}$50$}}%
      \put(814,1576){\makebox(0,0)[r]{\strut{}$100$}}%
      \put(814,1857){\makebox(0,0)[r]{\strut{}$150$}}%
      \put(814,2138){\makebox(0,0)[r]{\strut{}$200$}}%
      \put(814,2419){\makebox(0,0)[r]{\strut{}$250$}}%
      \put(814,2701){\makebox(0,0)[r]{\strut{}$300$}}%
      \put(814,2982){\makebox(0,0)[r]{\strut{}$350$}}%
      \put(814,3263){\makebox(0,0)[r]{\strut{}$400$}}%
      \put(946,818){\rotatebox{-45}{\makebox(0,0)[l]{\strut{}12/26}}}%
      \put(1307,818){\rotatebox{-45}{\makebox(0,0)[l]{\strut{}01/09}}}%
      \put(1669,818){\rotatebox{-45}{\makebox(0,0)[l]{\strut{}01/23}}}%
      \put(2030,818){\rotatebox{-45}{\makebox(0,0)[l]{\strut{}02/06}}}%
      \put(2391,818){\rotatebox{-45}{\makebox(0,0)[l]{\strut{}02/20}}}%
      \put(2752,818){\rotatebox{-45}{\makebox(0,0)[l]{\strut{}03/05}}}%
      \put(3114,818){\rotatebox{-45}{\makebox(0,0)[l]{\strut{}03/19}}}%
      \put(3475,818){\rotatebox{-45}{\makebox(0,0)[l]{\strut{}04/02}}}%
      \put(3607,1013){\makebox(0,0)[l]{\strut{}$0$}}%
      \put(3607,1294){\makebox(0,0)[l]{\strut{}$50$}}%
      \put(3607,1576){\makebox(0,0)[l]{\strut{}$100$}}%
      \put(3607,1857){\makebox(0,0)[l]{\strut{}$150$}}%
      \put(3607,2138){\makebox(0,0)[l]{\strut{}$200$}}%
      \put(3607,2419){\makebox(0,0)[l]{\strut{}$250$}}%
      \put(3607,2701){\makebox(0,0)[l]{\strut{}$300$}}%
      \put(3607,2982){\makebox(0,0)[l]{\strut{}$350$}}%
      \put(3607,3263){\makebox(0,0)[l]{\strut{}$400$}}%
    }%
    \gplgaddtomacro\gplfronttext{%
      \csname LTb\endcsname
      \put(209,2138){\rotatebox{-270}{\makebox(0,0){\strut{}Confirmed Cases}}}%
      \put(4245,2138){\rotatebox{-270}{\makebox(0,0){\strut{}Article Count}}}%
      \put(2210,154){\makebox(0,0){\strut{}Day and Month (2020)}}%
      \csname LTb\endcsname
      \put(2266,3090){\makebox(0,0)[r]{\strut{}Confirmed}}%
      \csname LTb\endcsname
      \put(2266,2870){\makebox(0,0)[r]{\strut{}Article}}%
    }%
    \gplbacktext
    \put(0,0){\includegraphics{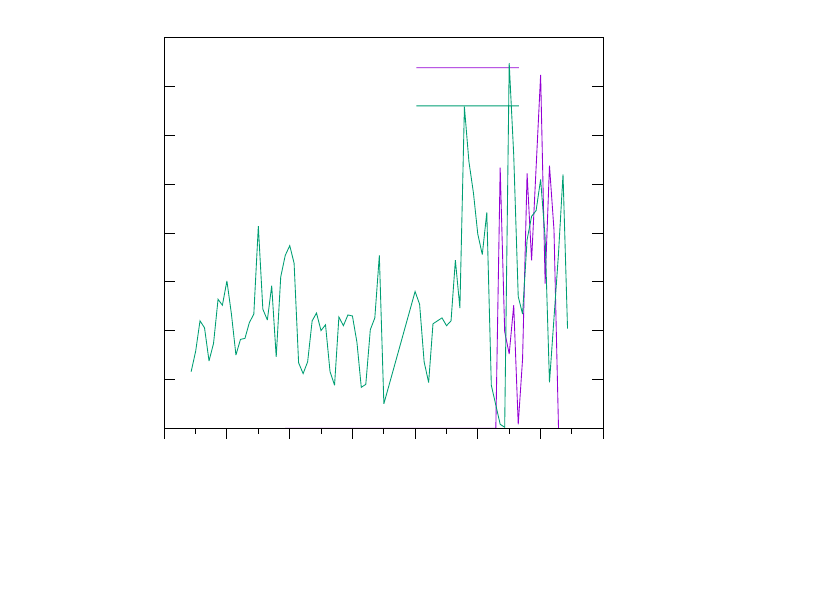}}%
    \gplfronttext
  \end{picture}%
\endgroup